\documentclass{article}

\usepackage[margin=2cm]{geometry}
\usepackage{mathtools, amsmath, amssymb, bm, siunitx}
\usepackage{graphicx,epstopdf}
\usepackage{colortbl}

\usepackage[style=numeric,uniquename=false,uniquelist=false,maxcitenames=1,maxbibnames=99,url=false,hyperref=true, sorting=ynt, isbn=false, eprint=false, giveninits=true,doi=true, backend=biber]{biblatex}
\usepackage{doi}
\DefineBibliographyStrings{english}{%
	andothers = {\em et\addabbrvspace al\adddot}
}
\DeclareNameAlias{sortname}{last-first}

\usepackage[font=small,labelfont=bf]{caption}
\usepackage{subcaption}
\usepackage{adjustbox}
\usepackage{booktabs}
\usepackage{xspace}
\usepackage{authblk}

\usepackage{hyperref}
\hypersetup{
	linkcolor  = black,
	citecolor  = red,
	urlcolor   = red,
	colorlinks = true,
}
\usepackage[capitalize]{cleveref}

\usepackage{tablefootnote}

\usepackage[automake]{glossaries}
\makeglossaries
\newacronym[\glslongpluralkey={Degrees of Freedom}]{dof}{DOF}{Degree of Freedom}
\newcommand{\dof}{\gls{dof}\xspace}
\newcommand{\dofs}{\glspl{dof}\xspace}

\newacronym[\glslongpluralkey={Degrees of Constraint}]{doc}{DOC}{Degree of Constraint}

\newcommand{\docs}{\glspl{doc}\xspace}

\newacronym[\glslongpluralkey={Finite element analyses}]{fea}{FEA}{Finite Element Analysis}
\newcommand{\fea}{\gls{fea}\xspace}

\newacronym{am}{AM}{Additive Manufacturing}

\newacronym{fact}{FACT}{Freedom And Constraint Topology}

\newacronym{topopt}{TO}{Topology optimization}

\newglossaryentry{var}{name={\ensuremath{x}},description={design variable},sort=x}
\newglossaryentry{xb}{name={\ensuremath{\mathbf{x}}},description={array of design variables},sort=xb}

\newglossaryentry{nvar}{name={\ensuremath{n}},description={number of design variables},sort=n}

\newglossaryentry{xu}{name={\ensuremath{\overline{x}}},description={variable upper bound},sort=xu}
\newglossaryentry{xl}{name={\ensuremath{\tilde{x}}},description={variable lower bound},sort=xl}

\newglossaryentry{setS}{name={\ensuremath{\mathbb{S}}},description={set of test cases}, sort=S}

\newglossaryentry{setA}{name={\ensuremath{\mathbb{C}}},description={set of test cases}, sort=A}
\newcommand{\setA}{\gls{setA}\xspace}
\newglossaryentry{setB}{name={\ensuremath{\mathbb{F}}},description={set of test cases}, sort=B}
\newcommand{\setB}{\gls{setB}\xspace}
\newglossaryentry{setX}{name={\ensuremath{\mathbb{X}}},description={set of variables}, sort=X}

\newglossaryentry{setE}{name={\ensuremath{\mathbb{E}}},description={set of variables}, sort=E}

\newcommand*{\tran}{^{\mkern-1.5mu\mathsf{T}}}
\usepackage{xcolor}
\definecolor{lightgray}{rgb}{0.63, 0.63, 0.63}

\newcommand{\abf}[4][]{\ensuremath{\mathbf{#2}_{\text{#3#4}#1}}}

\title{A simple and versatile topology optimization formulation for flexure synthesis}
\author[]{Koppen, S.} 
\author[]{Langelaar, M.}
\author[]{van Keulen, F.}

\affil[]{Department of Precision and Microsystems Engineering, Faculty of Mechanical, Maritime and Materials Engineering, Delft University of Technology, Mekelweg 2, 2628 CD Delft, The Netherlands\\ s.koppen@tudelft.nl}
\date{\today}

\begin{document}
\twocolumn
\maketitle

\begin{abstract}
High-tech equipment critically relies on flexures for precise manipulation and measurement.
Through elastic deformation, flexures offer extreme position repeatability within a limited range of motion in their degrees of freedom, while constraining motion in the degrees of constraint.
Topology optimization proves a prospective tool for the design of short-stroke flexures, providing maximum design freedom and allowing for application-specific requirements.
State-of-the-art topology optimization formulations for flexure synthesis are subject to challenges like ease of use, versatility, implementation complexity, and computational cost, leaving a generally accepted formulation absent.
This study proposes a novel topology optimization formulation for the synthesis of short-stroke flexures uniquely based on strain energy measures under prescribed displacement scenarios.
The resulting self-adjoint optimization problem resembles great similarity to `classic' compliance minimization and inherits similar implementation simplicity, computational efficiency, and convergence properties.
Numerical examples demonstrate the versatility in flexure types and the extendability of additional design requirements.
The provided source code encourages the formulation to be explored and applied in academia and industry.
\end{abstract}


%

\section{Introduction}
\label{sec:intro}

\begin{figure}
	\centering
	\begin{subfigure}[t]{0.15\textheight}   
		\includegraphics[width=\textwidth]{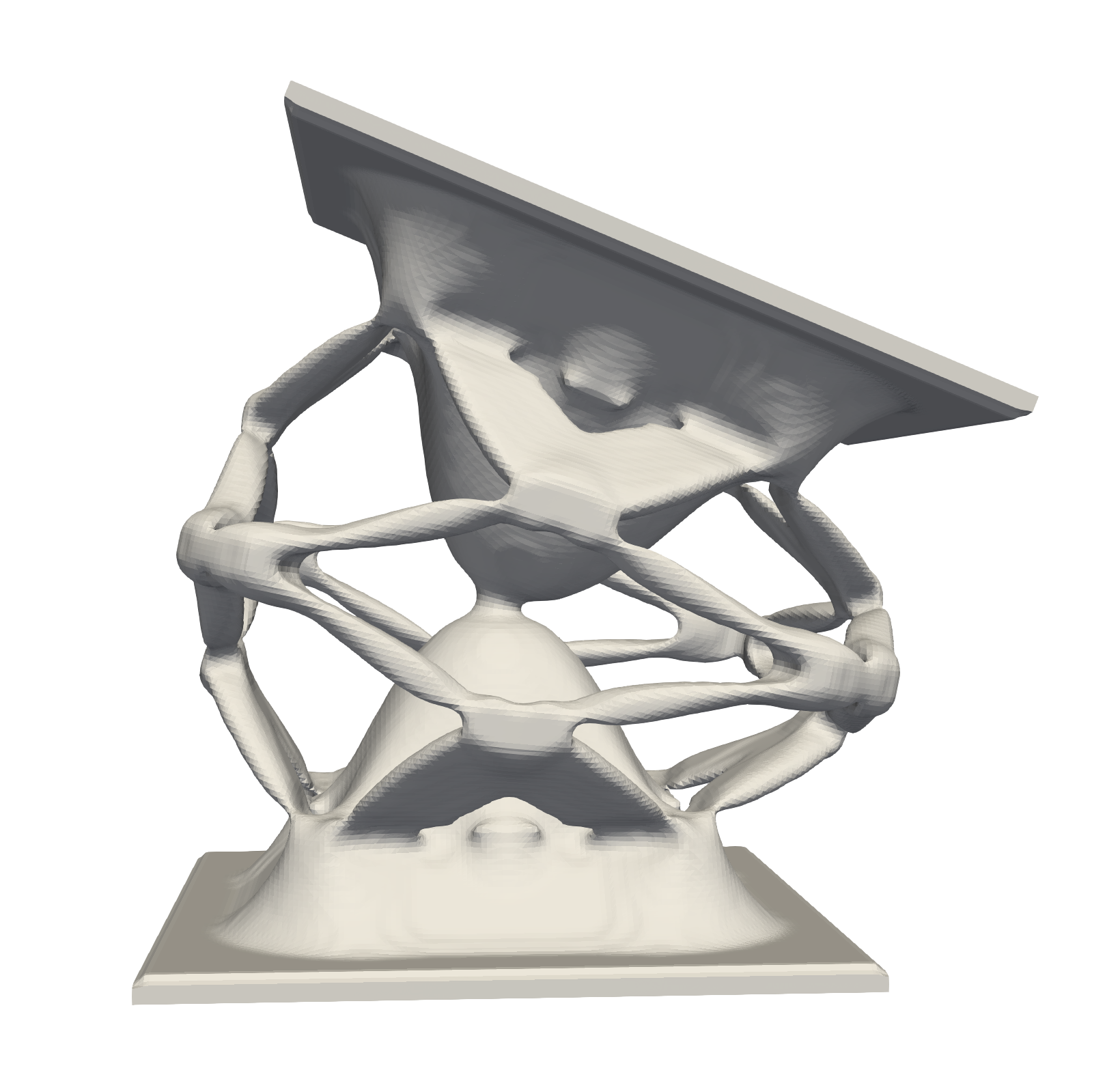}
	\end{subfigure}
	~
	\begin{subfigure}[t]{0.15\textheight}
		\includegraphics[width=\textwidth]{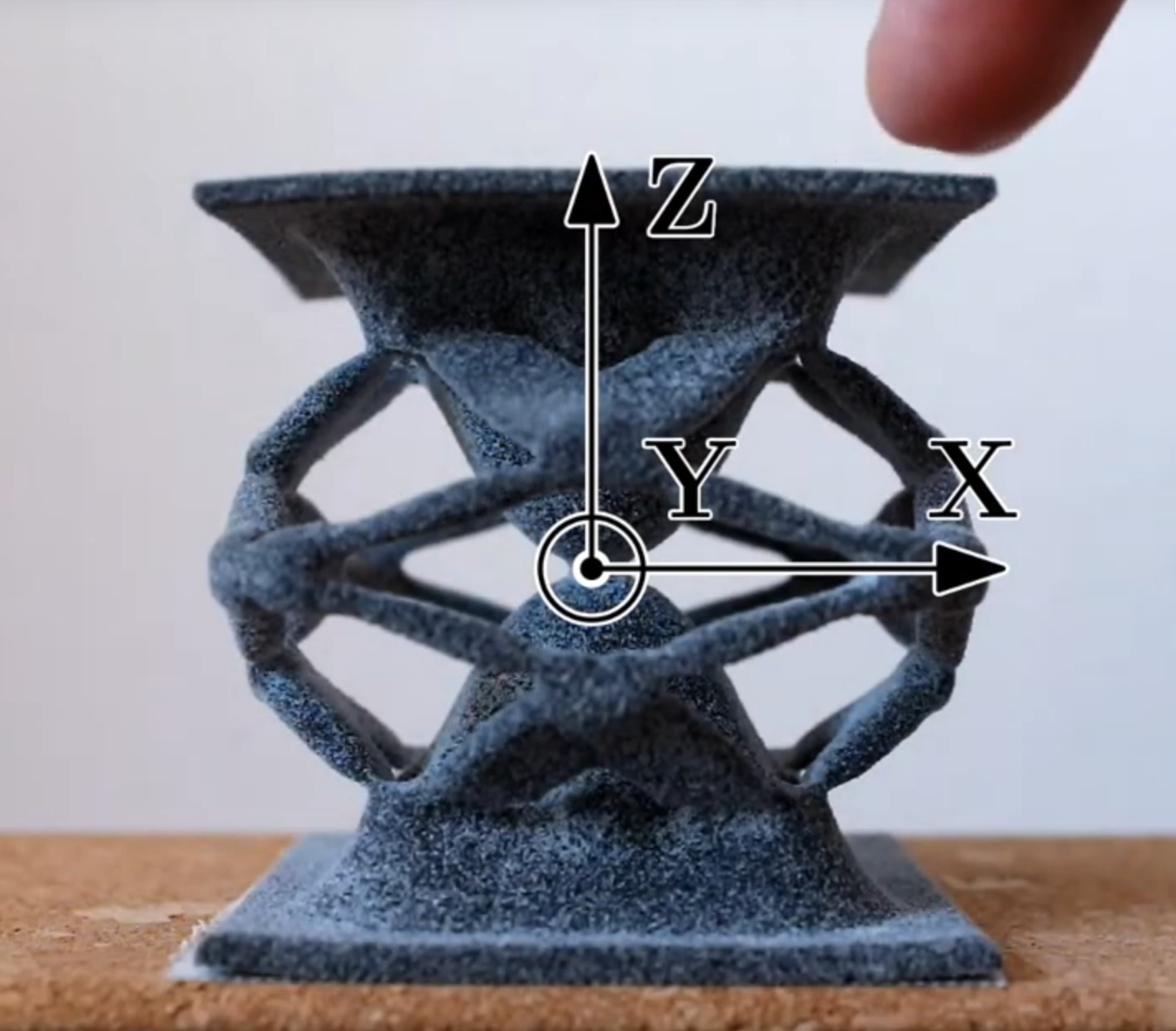}
	\end{subfigure}
	\caption{Computational design of a multi-axis flexure and its prototyped counterpart. The flexure is compliant in the rotations about the $x$ and $y$-axis 
		while stiff in the $x$, $y$ and $z$ translations 
		as well as rotation about the $z$-axis.}
	\label{fig:printed}
\end{figure}

A flexure is a monolithic compliant element that connects two or more (assumed) rigid links, allowing for selectively chosen movements.
Flexures are engineered to be compliant for specific relative rigid link movements, the mechanism \dofs, while stiff in other mechanism degrees, the mechanism \docs.
An example of a complex three-dimensional flexure is shown in \cref{fig:printed}.\footnote{A movie of the prototypes in motion can be found in the supplementary material attached to this paper. }

In contrast to conventional hinges, flexures achieve their range of motion through elastic deformation. The finite dimension and operation below a critical stress limit the attainable range of motion.
Due to the monolithic nature, flexures hardly require maintenance and have a long lifetime if used within the intended range of motion.
Due to the lack of friction and backlash, flexures have high repeatability in use. 
Given these advantages, flexures are commonly applied in precision applications such as positioning stages and optical mounts \autocite{Hu2017}. 
The present work is focused on so-called short-stroke flexures, for which---in contrast to large-stroke flexures---the assumptions of a linear stress-strain \emph{and} a linear strain-displacement relationship suffice.
Common single-axis flexures are (i) compliant revolute joints\footnote{Also called (flexural) hinge, flexure bearing or flexure pivot.}, such as notch hinges that allow relative rotation \textit{about} a single axis \autocite{Machekposhti2015}, and (ii) compliant prismatic joints\footnote{Also called translational (flexure) hinges.}, such as a pair of parallel leaf-springs, that allow relative motion \textit{along} a single axis. 
Common multi-axis flexures are, \textit{e.g.}, compliant cylindrical, universal, spherical and planar joints \autocite{Machekposhti2015}.
Complex flexures typically combine multiple primitive flexures as building blocks, thus enabling more complex kinematics \autocite{Gallego2009}.
Also, flexures can be classified by the degree of localization of the deformation, ranging between lumped (\textit{i.e.} highly localized) and distributed compliance \autocite{Yin2003}.


The primary design requirement of a short-stroke flexure is the relative stiffness between the mechanism DOFs and DOCs.
Secondary considerations are range of motion, axis drift, deformation and stress, fatigue, volume and mass, as well as the sensitivity of those aspects to, \textit{e.g.}, manufacturing errors.
The synthesis methods often used for rigid-body mechanisms, cannot straightforwardly be applied to compliant mechanisms. 
There is always mechanical stress involved in any motion, and the behaviour is dependent on the loading condition. 
This implies that kinematics (motion) and kinetics (load case) must be treated simultaneously. 
As a result, the concept of mechanism DOFs fades in compliant mechanisms, because they behave differently for any loading conditions \autocite{herder2017flexure}.
Furthermore, the complex deformation and motion behaviour of compliant mechanisms complicates both their accurate analytical modeling as well as purposeful design. Hence, the synthesis process is iterative, and often time-consuming \autocite{Linss2019}.

Systematic flexure synthesis methods rely on kinematic or building block approaches, such as rigid-body replacement techniques or the `freedom and constraint topology' method \autocite{Gallego2009}. However, these approaches do not exploit the full range of design possibilities.
The use of gradient-based structural optimization techniques to design flexures has gained increasing interest because of the possibility to design optimized flexures, satisfying application-specific requirements \autocite{Zhang2018b}. 
Topology Optimization (TO) in particular, allows for maximum design freedom, while requiring minimal designer input regarding the flexure concept \autocite{Bendsoe2004}.


Owing to the potential benefits of TO, academics, engineers and designers could benefit from a versatile, simple, easy to implement and use as well as computationally efficient TO method for short-stroke flexure design.
Multiple different TO problem formulations are previously proposed, see, \textit{e.g.}, \autocite{Hasse2009,Wang2009a,Zhu2014,Zhang2018b,Pinskier2019,Pinskier2020}.
\cref{sec:comparison} provides a short comparison of previously proposed TO problem formulations for flexure design and addresses the remaining challenges in the field.

To address the challenges, in \cref{sec:method} we propose a novel and intuitive topology optimization formulation to design flexures.
Staying close to the definition of a flexure, the basic idea is to maximize the stiffness of \emph{a priori} defined mechanism \docs, whilst imposing an upper bound on the stiffness of \emph{a priori} defined mechanism \dofs. 
Mechanism degree stiffnesses are evaluated via strain energies under prescribed movements of the rigid links.
In contrast to the traditional compliance minimization under applied load, maximization of strain energy under prescribed displacements results in maximization of corresponding stiffness and \textit{vice versa}.
The optimization problem is self-adjoint and obtaining the sensitivity information requires negligible computational effort.
The formulation is simple and generally applicable to design any type of flexure: single-axis, or multi-axis, planar and 3D. Independent of the formulation, implementation specific considerations are stated in \cref{sec:implementation}.

The basic formulation proposed in this work focuses on the primary design requirement, that is maximization of the relative stiffness between mechanism DOCs and mechanism DOFs. However, we will additionally demonstrate the ease and influence of taking into account stress considerations as well as manufacturing robustness in \cref{sec:results}, all within the limits of linear elasticity theory.
\Cref{sec:ror} outlines the code associated with this paper, with which the 2D results can be reproduced easily. The manuscript is completed with a reflection on the proposed formulation, limitations, recommendations for future work and conclusions.


\section{Comparison of existing formulations}
\label{sec:comparison}

Currently, a good comparison between different TO formulations to synthesize flexures is absent.
To compare different formulations, we define the following three quality criteria:
\begin{itemize}
	\item simplicity,
	\item versatility, and
	\item computational effort.
\end{itemize}
We define simplicity as the ease of understanding, implementation and use of the formulation. This includes the number of parameters required to define the optimization problem and the ease of assigning an appropriate value to those parameters.
Versatility is the applicability of the method to a wide range of uses \textit{e.g.} planar to three-dimensional or single-axis to multi-axis flexures.
The total computational effort to obtain an optimized design in a nested analysis and design process depends on the number of design iterations and the effort per design iteration. 
The number of design iterations is highly dependent on the ease of solving the resulting optimization problem and, thereto, the complexity (from an optimization point of view) of the optimization problem formulation.
The main contribution to the computational effort per design iteration is the number of analyses and their expense, such as a Finite Element Analysis (FEA). 
The effort of an analysis can be predominantly separated in the effort of the preconditioning/factorization (most expensive) and the iterative solve(s)/back-substitution(s), for iterative and direct solution approaches, respectively.

Below different approaches to flexure design using TO are discussed from the perspective of the aforementioned quality criteria. 
The aim of the discussion is to provide a concise overview of the field and build the argumentation for the present work. Thereto, in-depth review and/or comparison is out of the scope of this work. For a detailed description of the formulations the reader is referred to the relevant contributions, as presented in the first column of \cref{tab:perf}.
This table presents quantifiable measures of the quality criteria for each approach.
The following discussion adopts a categorization in kinetostatic and kinetoelastic formulations as proposed by \textcite{Wang2009a}.

\begin{table*}[t]
	\centering
	\caption{Topology optimization problem formulations for flexure design versus quantifiable measures of the quality criteria. The papers are ordered by year, ending with the present contribution. Versatility is expressed by the demonstrated range of applications (types of joints, single or multi-axis) and dimensionality (2D or 3D). The Parameters column denotes the \emph{minimum} number and type of parameters required to set up the formulation. Implementation includes notable features, such as the type of analyses and responses. The computational effort of a single design iteration is dominated by the effort of finite element and sensitivity analyses. The last column indicates, subsequently the number of (i) preconditioning/factorization steps, (ii) physical loads, and (iii) additional adjoint loads per design iteration. The sum of the loads indicates the number of iterative solves/back-substitutions required. For fair comparison all listed numbers of parameters and loads are for a single-axis flexure formulation. }
	\begin{tabular}{ccp{3cm}p{3.5cm}p{3cm}c}
		\toprule
		Paper & Dim & Versatility & Parameters & Implementation & Effort \\
		\midrule
		\arrayrulecolor{white}
		\cite{Hasse2009} & 2D & any single-axis joint & max volume \newline eigenmode
		& static condensation \newline orthogonalization \newline eigensystem analysis  & 1, 2, 0 \\
		\hline\\
		\cite{Wang2009a} & 2D	& any single or \newline multi-axis joint  & max volume \newline eigenmode & static condensation \newline eigensystem analysis & 1, 2, 0 \\
		\hline\\
		\cite{Zhu2014} 	&2D	& prismatic and \newline revolute joint & max volume \newline non-design domain size \newline max axis drift \newline spring stiffness & non-design domain  \newline exotic responses \newline additional spring & 1, 2, 1 \\
		\hline\\
		\cite{Pinskier2019} & 3D & leaf flexure & max volume \newline max strain energy & strain energy based & 1, 2, 0\\
		\hline\\
		\cite{Pinskier2020} & 2D & revolute joint & max volume \newline non-design domain size \newline max displacement & non-design domain \newline exotic responses & 1, 3, 3 \\
		\hline\\
		\arrayrulecolor{black}
		Present & 2D,3D & any single or \newline multi-axis joint & max strain energy & strain energy based & 1, 2, 0\\

		\bottomrule
		
	\end{tabular}
	
	\label{tab:perf}
\end{table*}

\subsubsection*{Kinetostatic formulations}
Naturally, TO problem formulations for flexure design find their origin in the field of compliant mechanism design.
Kinetostatics\footnote{Often referred to as `inverse dynamics'.} deals with the determination of forces that act upon the elements of a mechanism, given the mechanical system acts as a \emph{static} construction \autocite{burns1968kinetostatic}.
The so-called kinetostatic formulations, in one form or another, simultaneously aim to maximize the energy transmission between the input and output ports and mechanism's structural stiffness \autocite{Wang2009a}.
The mechanism performance is generally quantified using the concepts of mechanical advantage, geometric advantage, mechanical efficiency, flexibility-stiffness or mutual potential energy \autocite{Cao2013}.
%
Although there is no single universally accepted formulation, it has been shown that these formulations produce almost similar topologies for the optimized compliant mechanisms \autocite{Deepak2009}, \textit{viz.} these topologies tend to emulate their rigid-body counterpart \autocite{Wang2009a}.

Derived from these kinetostatic formulations for compliant mechanism design, \textcite{Zhu2014} and \textcite{Pinskier2020} independently proposed straightforward approaches for designing planar single-axis (prismatic and/or revolute) joints, taking into account axis drift.
The method of \textcite{Zhu2014} has been, among others, extended to account for geometric nonlinearity \autocite{Liu2015} and stress constraints \autocite{Liu2017b}.

\textcite{Pinskier2019} additionally proposed a simple and intuitive TO formulation aimed at the synthesis of leaf-springs using only strain-based measures. 
As a result, the formulation is simple and computationally efficient.
\textcite{Hu2017} applied TO to design 3D flexures for mounting the primary mirror of a space telescope by minimization of the gravity-induced surface shape error.

\subsubsection*{Kinetoelastic formulations}
As opposed to kinetostatic formulations, kinetoelastic formulations consider the mechanism's kinematic functions as an integral part of the \emph{elastic} properties of the continuum structure and seek to find compliant mechanisms with desirable intrinsic properties \autocite{Wang2009a}.
This is, thus far, accomplished by shaping the mechanism stiffness matrix entries.
The mechanism stiffness matrix is obtained by static condensation of the global stiffness matrix to a small set of nodal displacements
that can describe the mechanism degrees \autocite{Guyan1965,Irons1965}.
The formulation was effectively applied to the design of planar prismatic joints \autocite{Wang2009a}, and revolute joints \autocite{Li2019a}.

From a shape-morphing design philosophy, \textcite{Hasse2009,Hasse2017} proposed a kinetoelastic formulation to design compliant mechanisms with \textit{selective compliance} by shaping the modal properties of the mechanism stiffness matrix (\textit{i.e.} eigenmodes and eigenvalues). 
Compliant mechanisms with selective compliance combine the advantages of both lumped and distributed compliance, that is reduced stress concentrations and a distributed deformation pattern, while preserving defined kinematics \autocite{Hasse2017}. 

The kinetoelastic formulations of \textcite{Wang2009a,Hasse2009} use static condensation to obtain the mechanism stiffness matrix. 
This procedure requires an expensive analysis which scales with the number of nodal displacements required to describe the mechanism degrees.
Thereto, this is highly efficient for problems like single-input-single-output compliant mechanisms, for which the mechanism degree can, generally, be described using only two nodal displacements.
However, for the aforementioned problem formulations, a vast number of nodal displacements are required to describe the mechanism DOFs and DOCs.
As a result, applying static condensation (without further adaptation) to flexure design would generally require substantial high computational effort.

%

Despite the attention devoted to TO of flexures, the previously proposed formulations have disadvantages and pose challenges, see also \cref{tab:perf}.
The kinetostatic formulations are straightforward but tend to be specific for a small set of flexures \autocite{Zhu2014,Pinskier2019,Pinskier2020}.
In contrast, the kinetoelastic formulations are versatile, however are generally more complex to implement \autocite{Hasse2009, Wang2009a}. 
Several formulations include responses that depend highly nonlinear on the nodal displacements \autocite{Wang2009a,Pinskier2020} or make use of artificial stiffness and additional user-defined parameters \autocite{Zhu2014}, that can make application difficult.
Some show inferior convergence properties (many iterations or oscillatory behaviour) and/or deliver non-binary (and hence non-manufacturable) topologies due to absence of conflicting requirements \autocite{Zhu2014,Pinskier2020}.
Finally, some formulations require substantial computational effort, which makes application of the method unpracticable.
To conclude; none of the previously proposed formulations is simple to understand, implement and use \emph{as well as} versatile \emph{and} computationally efficient.

\section{Method}
\label{sec:method}


\begin{figure*}
	\centering
	\includegraphics[width=\textwidth]{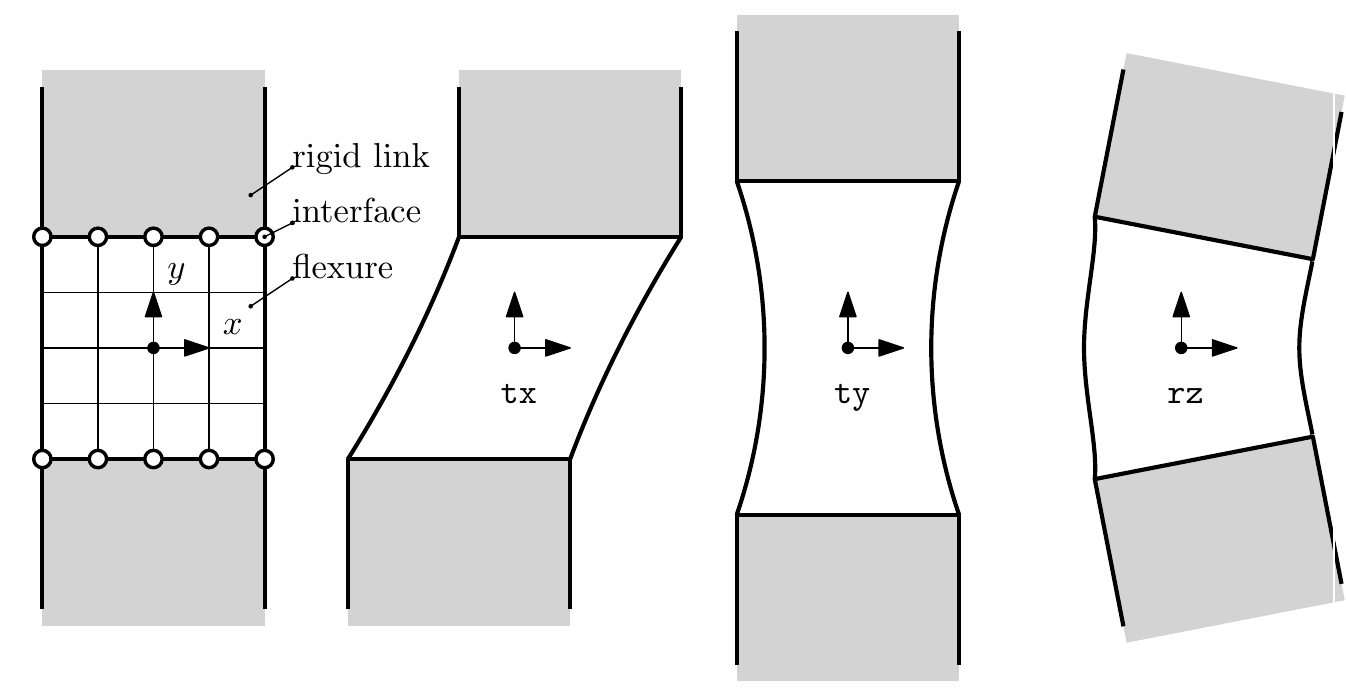}
	\caption{On the left: two rigid links (gray) connected via a flexure (white). The flexure geometry is discretized using $\texttt{nelx} \times \texttt{nely}$ finite elements. The interface nodes used to prescribe the mechanism degrees are here denoted by circles ($\circ$). Those nodal displacements are used to prescribe different mechanism degrees. On the right: three different mechanism degrees commonly used in 2D flexure design; relative translation along the x-axis (\texttt{tx}) and y-axis (\texttt{ty}) and rotation about the z-axis (\texttt{rz}).}
	\label{fig:loadcases}
\end{figure*}

Consider a structure within a bounded domain $\Omega$, made of an isotropic linear elastic material. 
For simplicity of explanation, we discretize the domain in a structured grid of $N$ finite elements (\texttt{nelx} $\times$ \texttt{nely})\footnote{Typewriter font is used to indicate the names as used in the code associated with this work to replicate the results, see \cref{sec:ror}.} with a total of $n$ nodal displacements, as sketched in \cref{fig:loadcases}.
Let us define a set $\mathbb{D}$, consisting of unique mechanism degrees (DOFs and DOCs). 
For example, consider the set $\mathbb{D} = \left\{\texttt{tx}, \texttt{ty}, \texttt{rz}\right\}$, in line with the 2D problem depicted in \cref{fig:loadcases}. 
These mechanism degrees define prescribed nodal displacements at the interfaces between rigid link and flexure (\textit{e.g.} a unit displacement in $x$-direction between top and bottom interfaces for mechanism degree \texttt{tx}).
The assumption that these interfaces are rigid is valid if the links can be considered much stiffer compared to the flexure.
As such, the mechanism degrees correspond to the relative rigid body motions of the interfaces.
 
We define subset $\setA \subset \mathbb{D}$ that contains the mechanism \docs, and subset $\setB = \mathbb{D} \setminus \setA$ that contains the mechanism \dofs.
In line with the primary design requirement for short-stroke flexures, we aim to maximize the stiffness of the mechanism  DOCs in $\setA$, while constraining the maximum stiffness of the mechanism DOFs in $\setB$.
The present work uses the strain energy of the mechanism degrees as a measure for stiffness.
The proposed constrained nonlinear optimization problem formulation for flexure synthesis now simply reads
\begin{equation}
\mathcal{P} =\left\{
\begin{aligned}
\underset{\mathbf{x}}{\text{minimize}} && \quad & -f\left[\mathcal{E}_i\left[\mathbf{x}\right]\right], & i \in \mathbb{C}\\
\text{subject to}&&&  \mathcal{E}_j\left[\mathbf{x}\right] \leq \overline{\mathcal{E}}_j, &j \in \mathbb{F}\\
&&& \mathbf{x} \in \mathcal{X}^N
\end{aligned}
\right. ,
\label{eq:Pbasic}
\end{equation}
where the dimensionless objective $f \in \mathbb{R}^{+}$ is a monotonically increasing function of strain energies $\mathcal{E}_i \in \mathbb{R}^{+}$ and $\overline{\mathcal{E}}_j \in \mathbb{R}^{+}$ is the maximum allowable strain energy of mechanism degree $j$. The topology is described by $N$ continuous differentiable design variables $\mathbf{x}$ with its components in $\mathcal{X} := \left\{x \in \mathbb{R} \ | \ 0 < x \leq 1 \right\}$.


The strain energy of mechanism degree $i$ in a discretized setting is defined as
\begin{equation}
	\label{eq:sed}
	\mathcal{E}_i\left[\mathbf{x}\right] = \frac{1}{2} \abf[i]{u}{}{}\cdot\abf[]{K}{}{}\left[\mathbf{x}\right] \abf[i]{u}{}{}, 
\end{equation}
where $\mathbf{K}\left[\mathbf{x}\right] \in \mathbb{R}^{n \times n}$ is the design dependent symmetric stiffness matrix and $\mathbf{u}_i \in \mathbb{R}^{n}$ contains the nodal displacements of mechanism degree $i$.
These nodal displacements are obtained by analysis of the structural behaviour, described by $n$ linear governing equations
\begin{equation}
\label{eq:kuf}
\mathbf{K}\left[\mathbf{x}\right]\mathbf{u}_i = \mathbf{f}_i, \quad \forall ~ i \in \mathbb{D},
\end{equation}
where $\mathbf{f}_i \in \mathbb{R}^{n}$ are the nodal loads of mechanism degree $i$. 
To calculate $\mathbf{u}_i$, we partition \cref{eq:kuf} as
\begin{equation}\label{eq:eqofmotion}
\begin{bmatrix}
\abf[]{K}{f}{f} & \abf[]{K}{f}{p}\\
\abf[]{K}{f}{p}\tran & \abf[]{K}{p}{p}
\end{bmatrix}
\begin{bmatrix}
\abf[,i]{u}{f}{}\\
\abf[,i]{u}{p}{}
\end{bmatrix}
=
\begin{bmatrix}
\abf[,i]{f}{f}{}\\
\abf[,i]{f}{p}{}
\end{bmatrix},
\end{equation}
where \abf[,i]{u}{f}{} are the free nodal displacements, \abf[,i]{u}{p}{} the prescribed nodal displacements, \abf[,i]{f}{f}{} the applied nodal loads and \abf[,i]{f}{p}{} the nodal reaction loads of mechanism degree $i$. 
As mentioned, the mechanism degrees are defined purely in terms of prescribed nodal displacements at the interfaces, without additional applied loads. Hence,  the applied loads $\abf[]{f}{f}{} = \mathbf{0}$ in all cases. The solutions to \cref{eq:eqofmotion}, \abf[,i]{u}{f}{}, can be obtained by solving the system of linear equations
\begin{equation}
\label{eq:y2}
\abf[]{K}{f}{f} \abf[,i]{u}{f}{} = - \abf[]{K}{f}{p} \abf[,i]{u}{p}{}, \quad \forall ~ i \in \mathbb{D}.
\end{equation}


The allowable strain energies of the mechanism DOFs are the primary design requirement. These energies are generally known from system requirements, or can be derived from the required stroke for a given maximum actuation force or \emph{vice versa} from the required actuation force for a given stroke.

\subsubsection*{Sensitivity analysis}

TO generally requires the consecutive calculation of structural responses (objectives or constraints) and their sensitivity to the design variables.
Both generally involve one or multiple computationally expensive FEA. 
For specific optimization responses---for example strain-energy---the problem becomes so-called `self-adjoint' \autocite{Rozvany1993}.
In self-adjoint problems, the loading terms of the analyses required to obtain the structural response and sensitivity information are linearly dependent.
As a result, the computational cost of the sensitivity analysis reduces dramatically. 
Note that this advantage is only applicable to the linear case, which is the focus of this study.
All responses in $\mathcal{P}$ are self-adjoint. As a result, the sensitivities can be calculated based on available information. In addition, the sensitivities are separable, \textit{i.e.} each design variable contributes solely via its elemental strain energy. 
Thereto, one may write
\begin{equation}
\frac{\text{d}\mathcal{E}_i}{\text{d}x_j} = \gamma_{i,j}\left[x_j\right] \varepsilon_{i,j},
\end{equation}
with $\varepsilon_{i,j} \in \mathbb{R}^+$ and $\gamma_{i,j} \in \mathbb{R}$ the elemental strain energy and multiplication factor of element $j$ due to degree $i$.
The interpretation and derivation of $\gamma_{i,j}$ is further explained in \cref{sec:implementation}.
Since element strain energies (or elemental stiffness matrices in combination with the nodal displacements) are common output data in commercial finite element analysis software, the sensitivity analysis is straightforward to implement, even when using software packages that do not already provide sensitivity information.



Note that, due to the generality of the method, the problem formulation can include one or multiple \docs in the objective while constraining the stiffness of one or multiple \dofs. 
As such, the formulation can be used to design both single-axis and multi-axis flexures. 
Although $\mathcal{P}$ is relatively simple and only involves strain energy contributions from the considered degrees, it is effective in generating many types of flexures, as will be shown in \cref{sec:results}. 
Remarkably, this simple formulation for flexure synthesis has not been reported before to the best knowledge of the authors.

\section{Implementation}
\label{sec:implementation}

Independent of the problem formulation as presented, the user has to consider, select and implement a variety of methods to effectively use the formulation in a TO setting. 
Without loss of generality, the following aids in the consideration and implementation of design parametrization, filtering, material interpolation, response formulation and gradient-based optimization. 
All numerical examples employ the implementation choices described here.
The default constants used in the examples, as implemented in the attached code, are listed in \cref{tab:constants}.

\begin{table}[t]
		\caption{Constant parameters and assigned values.}
	\centering
	\begin{tabular}{c|l|l}
		\toprule
		Symbol & Description & Value\\
		\midrule
		$\varepsilon$ & Stiffness ratio & \SI{e-6}{}\\
		$\nu$ & Poisson ratio & \SI{0.3}{}\\
		$p$ & SIMP penalty & \SI{3.0}{} \\
		$r$ & filter radius (no. elements) & \SI{2.0}{} \\
		$\epsilon$ & maximum design change & \SI{e-3}{}  \\
		$x^0$ & homogeneous initial design & \SI{0.5}{}\\
		\bottomrule
	\end{tabular}
	\label{tab:constants}
\end{table}

For the \fea, we opt for standard 4-node quadrilateral (2D) and 8-node hexahedral elements (3D) in structured meshes.
The design domain is parametrized by assignment of a design variable $x_i \in \mathcal{X}$ to each finite element $i$, which allows for local control of the material properties \autocite{Bendsoe2004}.

It is generally recognized that both final design and performance are sensitive to the initial design $\mathbf{x}^{\{0\}}$.
This is especially the case for compliant mechanisms, and thus also for flexure optimization \autocite{Sigmund1998,Chen2017}.
We consider this influence out of the scope of this paper and thereto opt for the commonly used homogeneous initial design.

To eliminate modeling artifacts, the design variable field is generally blurred as to obtain the filtered field $\tilde{\mathbf{x}} \in \mathcal{X}^N$ using a linear filtering operation $\mathbf{H}\left[\mathbf{x}\right]: \mathcal{X}^N \rightarrow \mathcal{X}^N$ with relative filter radius $r \in \mathbb{R}^{+}$, see \textit{e.g.} \textcite{Bruns2001}. This operation is also accounted for in the sensitivity calculation, as described in the cited reference.

Asymmetric topologies resulting from problems with symmetric boundary conditions is, although not often explicitly reported, common and inevitable. 
The gradient-based optimizer solves many independent convex problems until a finite convergence criterion is met. 
As a result, round-off errors are inevitable. 
This leads, in most cases, to divergence from the symmetric local optimum. 
One may easily enforce symmetry by linking design variables over one or multiple axes; either by creating a dependency or by averaging.

The Young's modulus of an element is related to the filtered design variable via a element-wise composite rule, that is
\begin{equation}
\frac{E_i\left[\tilde{x}_i\right]}{\overline{E}} = \varepsilon + \left(1 - \varepsilon\right) R\left[\tilde{x}_i\right],
\end{equation}
with $\overline{E}$ the material Young's modulus, $\varepsilon$ the relative stiffness between solid and void and $R$ the material interpolation function.
We apply the commonly used modified Solid Isotropic Material with Penalization (SIMP) interpolation function proposed by \textcite{Bendsoe1989}, that is
\begin{equation}
R\left[x\right] = x^p,
\end{equation}
with $p \in \mathbb{R}^{+}$ a user definable parameters. 
It is commonly known that this interpolation function increases the probability to obtain a 0/1 solution of a strain-based optimization problem.
Note that, as a result, the elemental multiplication factor is simply obtained via
\begin{equation}
\gamma_{i,j} := (1-\varepsilon)\frac{\partial R_i}{\partial x_j}.
\end{equation}


It is generally beneficial to scale the objective such that it holds a reasonable value (as compared to the constraints) \cite{Svanberg1987}.
We opt to normalize the strain energy of degree $i$ to its strain energy at the first optimization iteration, that is
\begin{equation}
\label{eq:alpha}
\alpha_i^{\{k\}}: = \frac{\mathcal{E}_i^{\{k\}}}{\mathcal{E}_i^{\{0\}}}
\end{equation}
with $\alpha$ the relative strain energy and $k$ the optimization iteration counter. 
Note that, as a result of \cref{eq:alpha}, the normalized strain energy $\alpha$ is a dimensionless positive scalar value by definition and $\alpha_i^{\{0\}} = 1$ for all $i$. 

In order to simultaneously maximize the stiffness of multiple \docs, the corresponding normalized strain energies are combined in a monotonically increasing function.
We opt here for a simple summation, that is the objective at iteration count $k$ yields
\begin{equation}
	f^{\{k\}}\left[\mathcal{E}_i\right] := \sum_i^{|\mathbb{C}|}\alpha_i^{\{k\}}\left[\mathcal{E}_i\right]
\end{equation}
with $|\mathbb{C}|$ the number of DOCs.
One might, in addition, add weight factors to the individual strain energy measures to control relative importance or opt for a smooth minimum function \autocite{Ma1994}. Note that, as a consequence of \cref{eq:alpha}, the magnitude of the prescribed displacements for different DOCs become irrelevant.

The gradient-based inequality-constrained nonlinear optimization problem $\mathcal{P}$ is solved in a nested analysis and design setting.
The design variables are iteratively updated by a sequential approximate optimization scheme, as is common in the topology optimization field. 
We use the Method of Moving Asymptotes (MMA) by \textcite{Svanberg1987}. The resulting convex sub-problems are solved using a primal-dual interior point method. The optimization is terminated when the maximum design change is smaller than $\epsilon$. 

This work includes a MATLAB code to design 2D single and multi-axis flexures, which is discussed in more detail in \cref{sec:ror}. 
We provide a briefly introduction here to allow the reader to understand how to replicate the results in upcoming sections.
The code is an extension of the commonly used \texttt{top71.m} code by \textcite{Andreassen2011}, and can be called using a similar syntax, that is
\begin{equation*}
\texttt{flexure(nelx, nely, doc, dof, emax)}
\end{equation*}
where \texttt{doc} and \texttt{dof} are lists of strings of the names of \docs and \dofs, respectively. Parameter \texttt{emax} is a list of maximum allowable strain energies corresponding to the \dofs in \texttt{dof}. 


\section{Numerical examples}
\label{sec:results}



In order to demonstrate the method proposed in \cref{sec:method}, we apply it to common problems for which results have been reported in literature. 
Thereto, we introduce a set $\mathbb{D} = \left\{\texttt{tx}, \texttt{ty}, \texttt{rz}\right\}$ consisting of the three rigid body degrees of the rigid links; two relative translations and rotation around the center of the flexure, see \cref{fig:loadcases}.
A sketch of the deformed structure resulting from the prescribed degrees for $\mathbf{x}^{\{0\}}$ are shown in \cref{fig:loadcases}.
Note that, without adjusting the formulation, any other set of unique degrees may be used.

\begin{figure*}[t]
	\centering
	\hfill
	\begin{subfigure}[t]{0.3\textwidth}
		\includegraphics[angle=90,width=\textwidth]{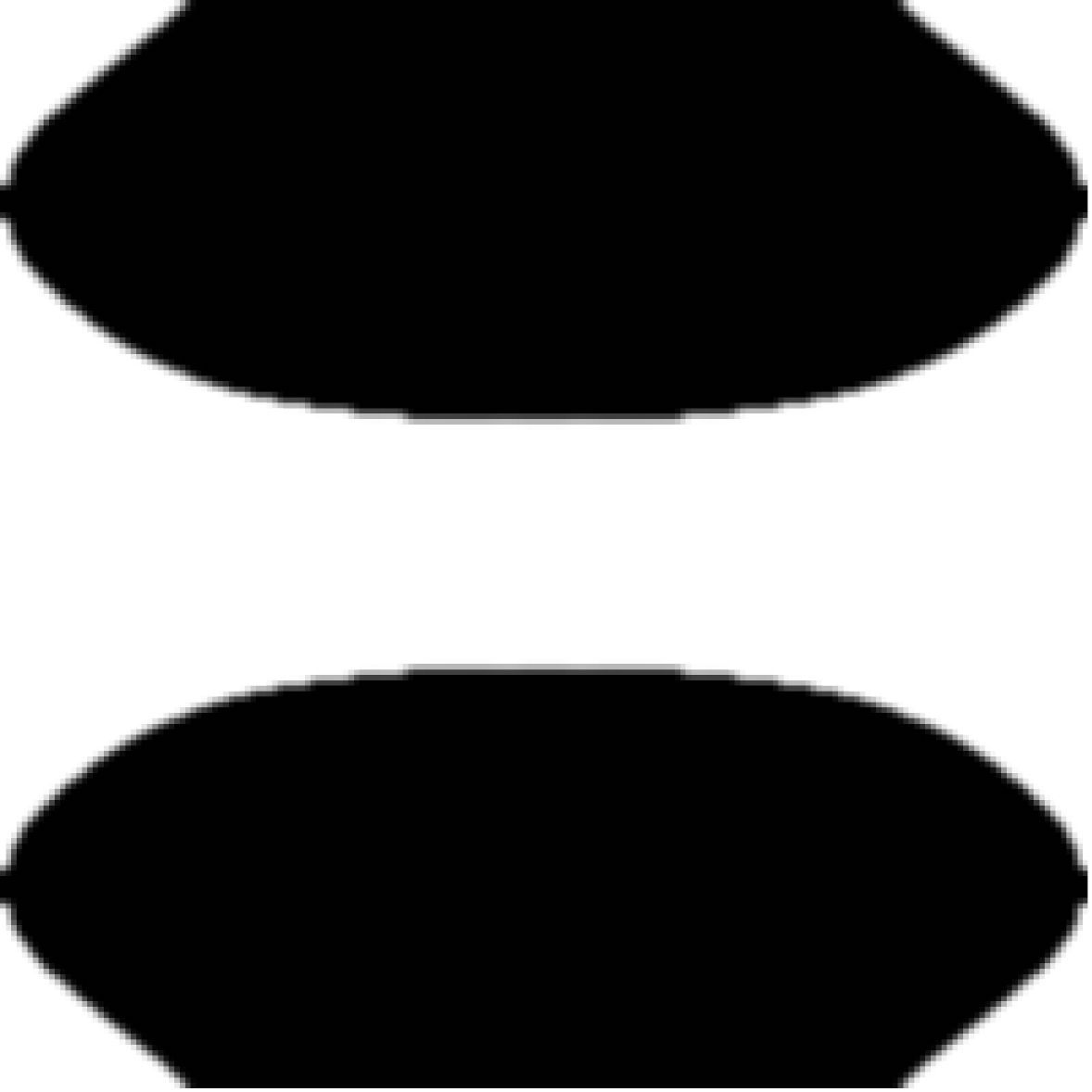}
		\caption{\texttt{doc=ty, dof=tx}}
		\label{fig:2dmaxtymintx}
	\end{subfigure}
	\hfill
	\begin{subfigure}[t]{0.3\textwidth}
		\centering
		\includegraphics[angle=90,width=\textwidth]{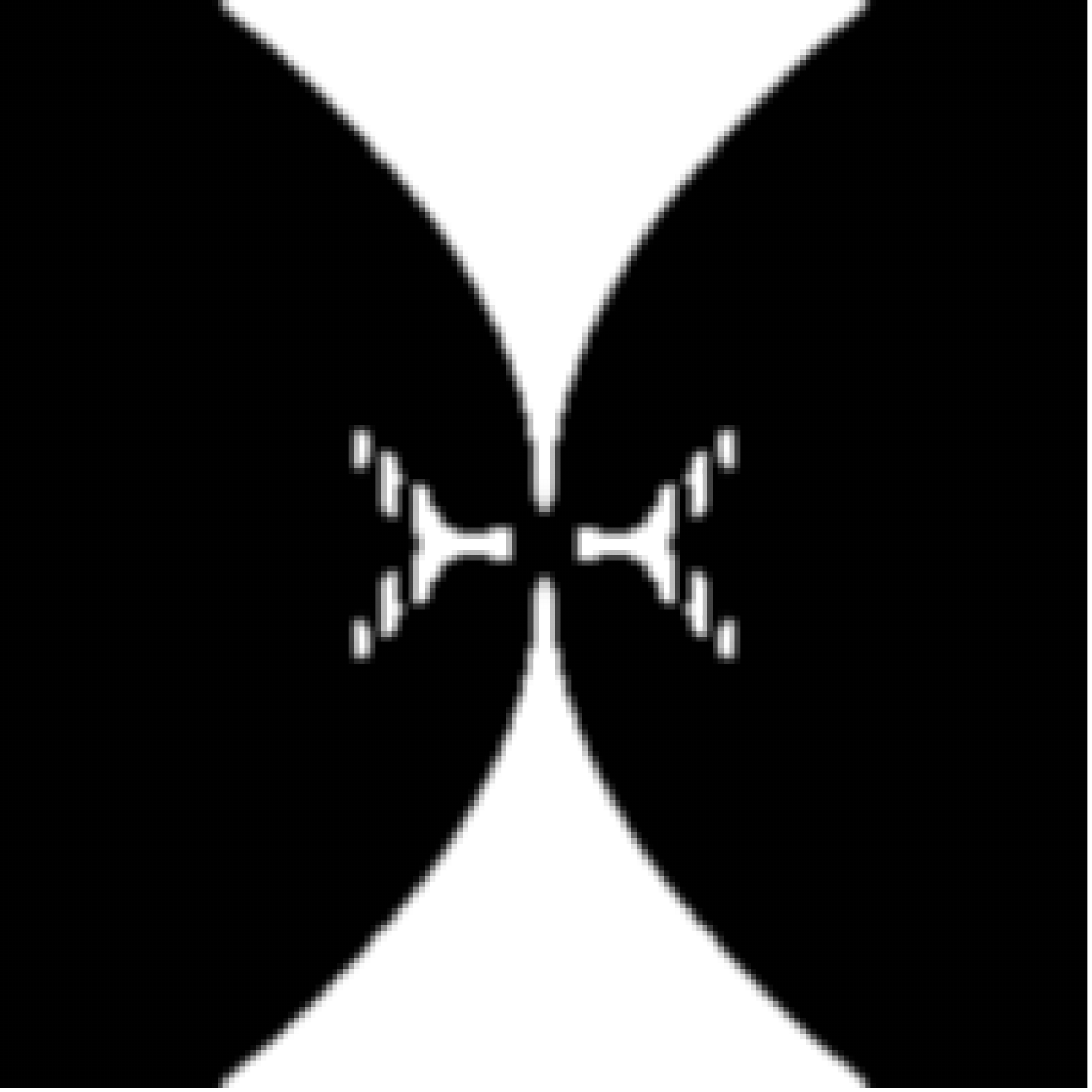}
		\caption{\texttt{doc=tx, dof=rz}}
		\label{fig:2dmaxtxminrz}
	\end{subfigure}
	\hfill
	\begin{subfigure}[t]{0.3\textwidth}
		\includegraphics[angle=90,width=\textwidth]{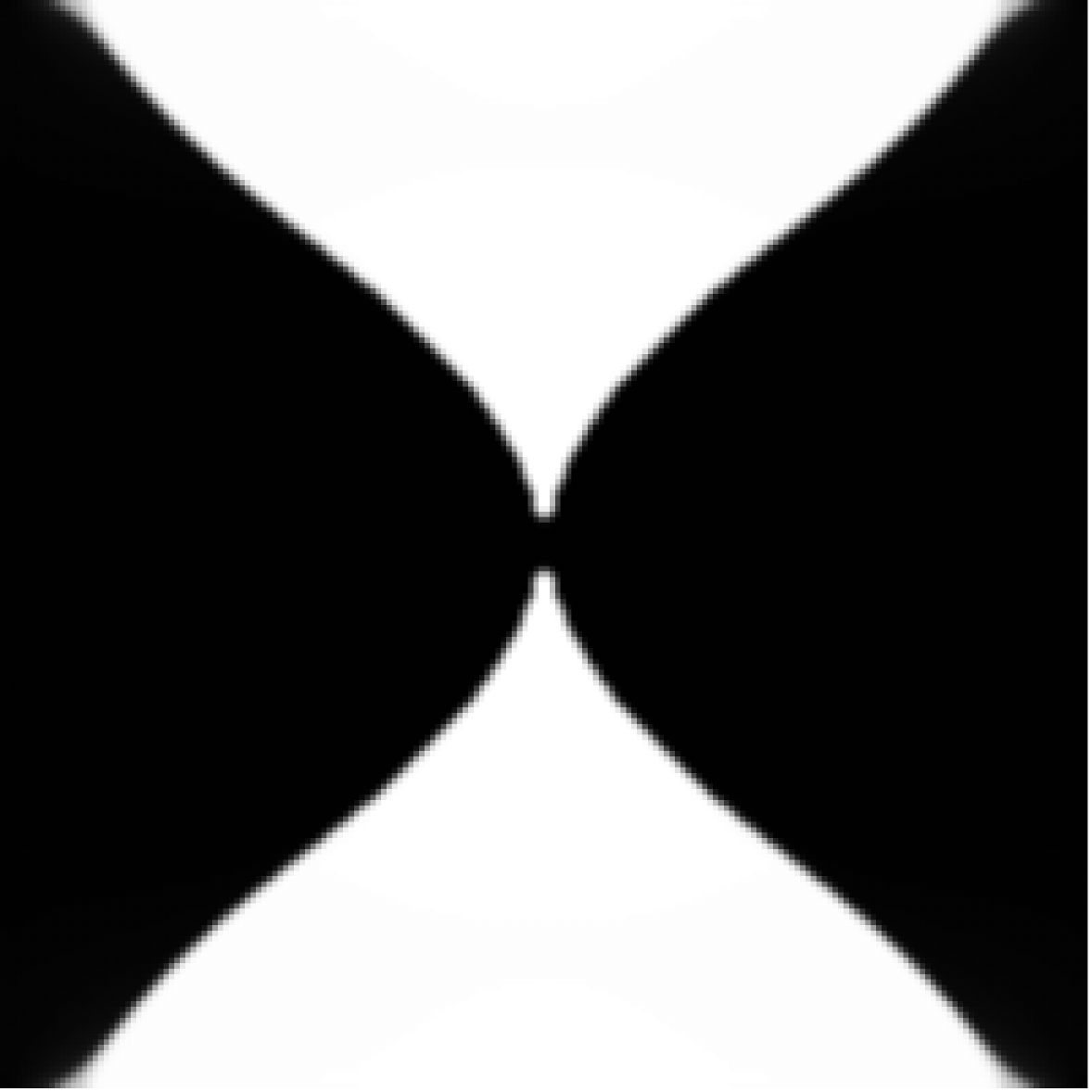}
		\caption{\texttt{doc=ty, dof=rz}}
		\label{fig:2dmaxtyminrz}
	\end{subfigure}
	\hfill
	\vskip\baselineskip
	\hfill
	\begin{subfigure}[t]{0.3\textwidth}
		\centering
		\includegraphics[angle=90,width=\textwidth]{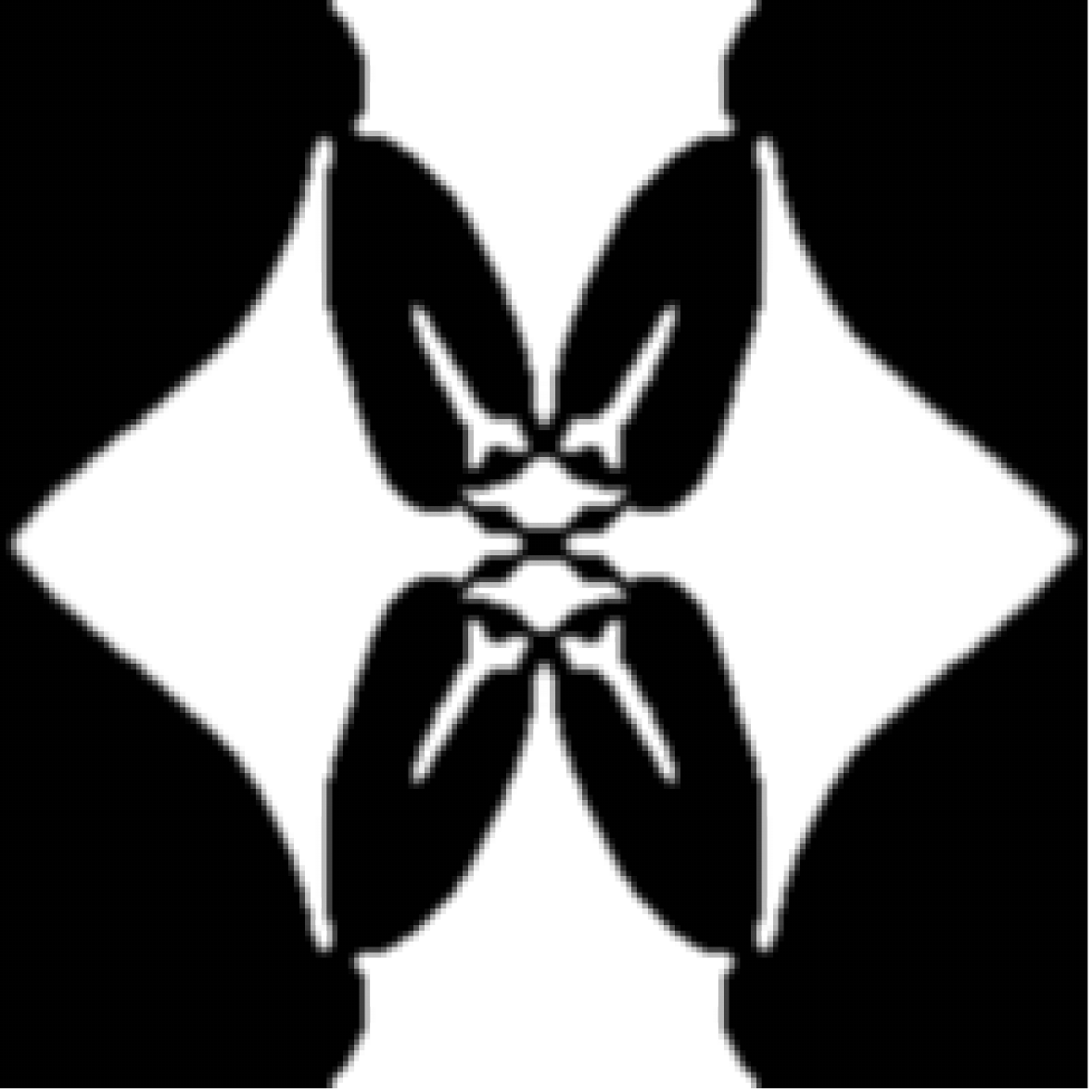}
		\caption{\texttt{doc=tx, dof=ty}}
		\label{fig:2dmaxtxminty}
	\end{subfigure}
	\hfill
	\begin{subfigure}[t]{0.3\textwidth}
		\centering
		\includegraphics[width=\textwidth]{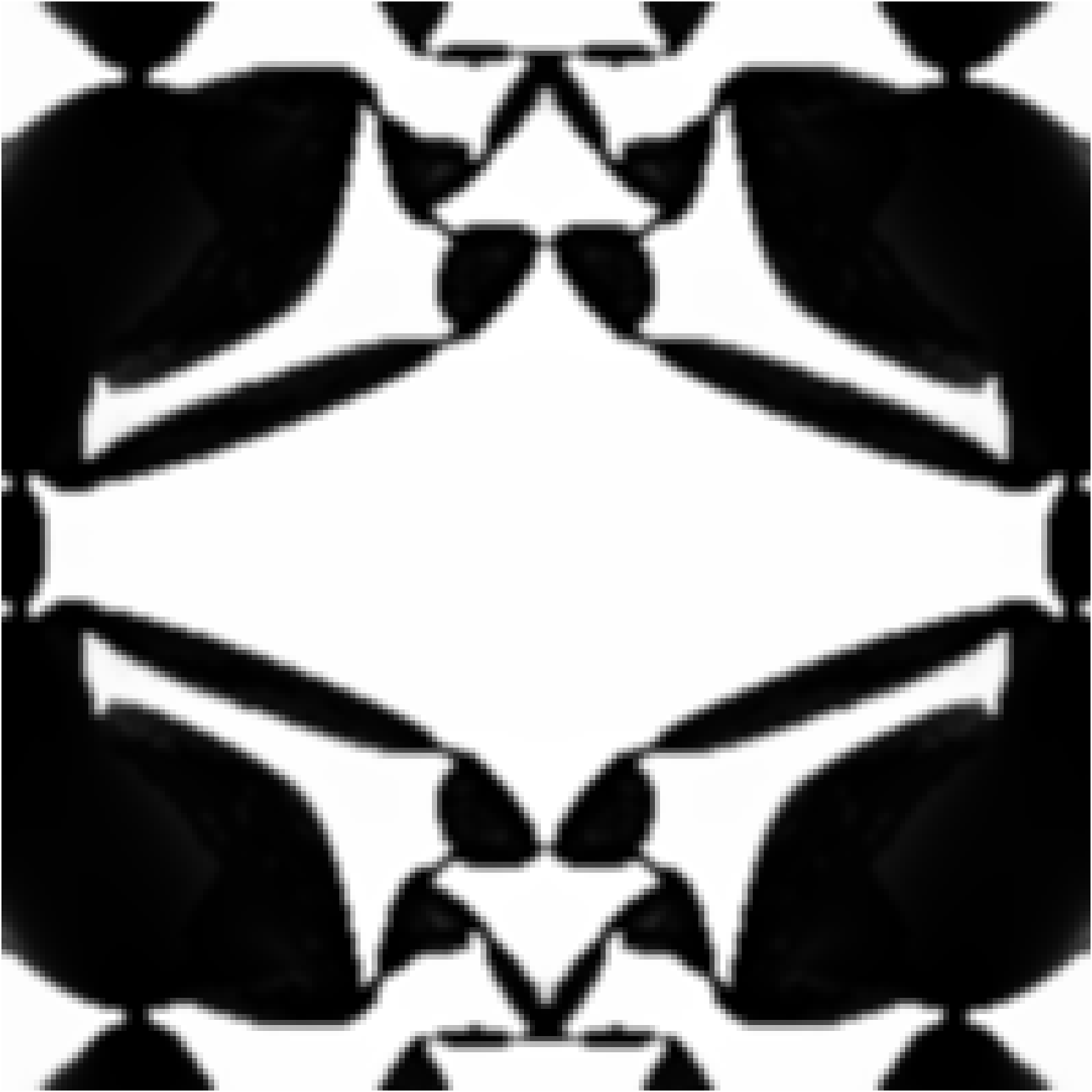}
		\caption{\texttt{doc=rz, dof=[tx,ty]}}
		\label{fig:2dmaxtxyminrz}
	\end{subfigure}
	\hfill
	\begin{subfigure}[t]{0.3\textwidth}
		\centering
		\includegraphics[width=\textwidth]{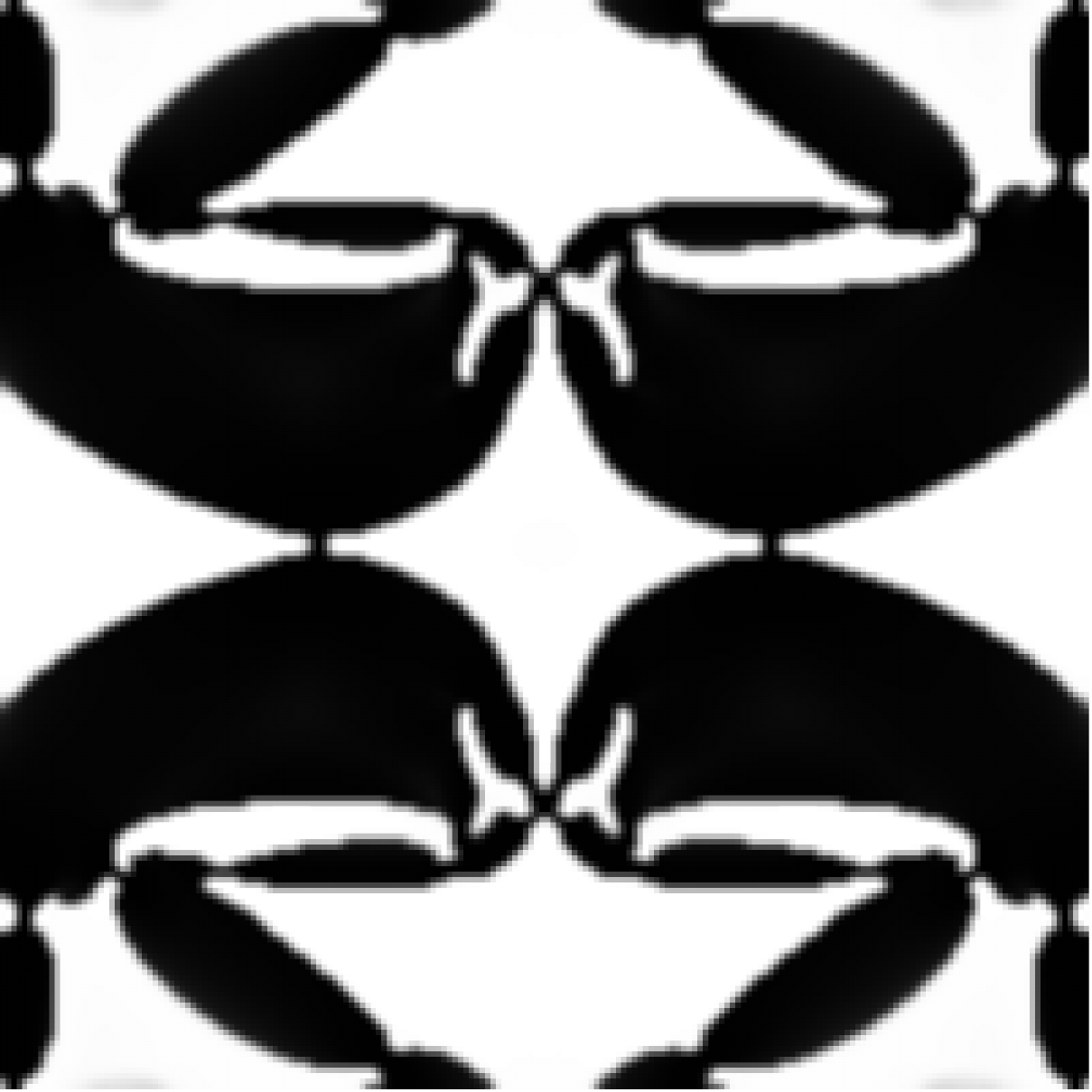}
		\caption{\texttt{doc=[ty,rz], dof=tx}}
		\label{fig:2dmaxtyrzmintx}
	\end{subfigure}
	\hfill
	\caption{Generated design for various varieties of problem formulations based on $\mathcal{P}$. This includes both single and multi-axis mechanisms for 2D topologies. These designs are generated by \texttt{flexure(200,200,doc,dof,emax)}, with \texttt{doc} and \texttt{dof} given in the subcaptions and the maximum strain energies in \texttt{emax} equal for all cases. These results can be replicated using the attached code. Note that we have omitted the string signature (\textit{e.g.} \texttt{"tx"}) here for simplicity.} 
	\label{fig:flexures2d}
\end{figure*}

\cref{fig:flexures2d} show the resulting topologies for a variety of planar design cases. Primitive topologies for the design of a compliant prismatic and revolute joint are shown in \cref{fig:2dmaxtymintx} and \cref{fig:2dmaxtyminrz} respectively.
The results are as expected and fully in accordance to the results obtained by both synthesis methods \autocite{Howell2001,Gallego2009} and topology optimization formulations \autocite{Wang2009a,Zhang2018b}.

\begin{figure}
	\centering
	\includegraphics[width=0.45\textwidth]{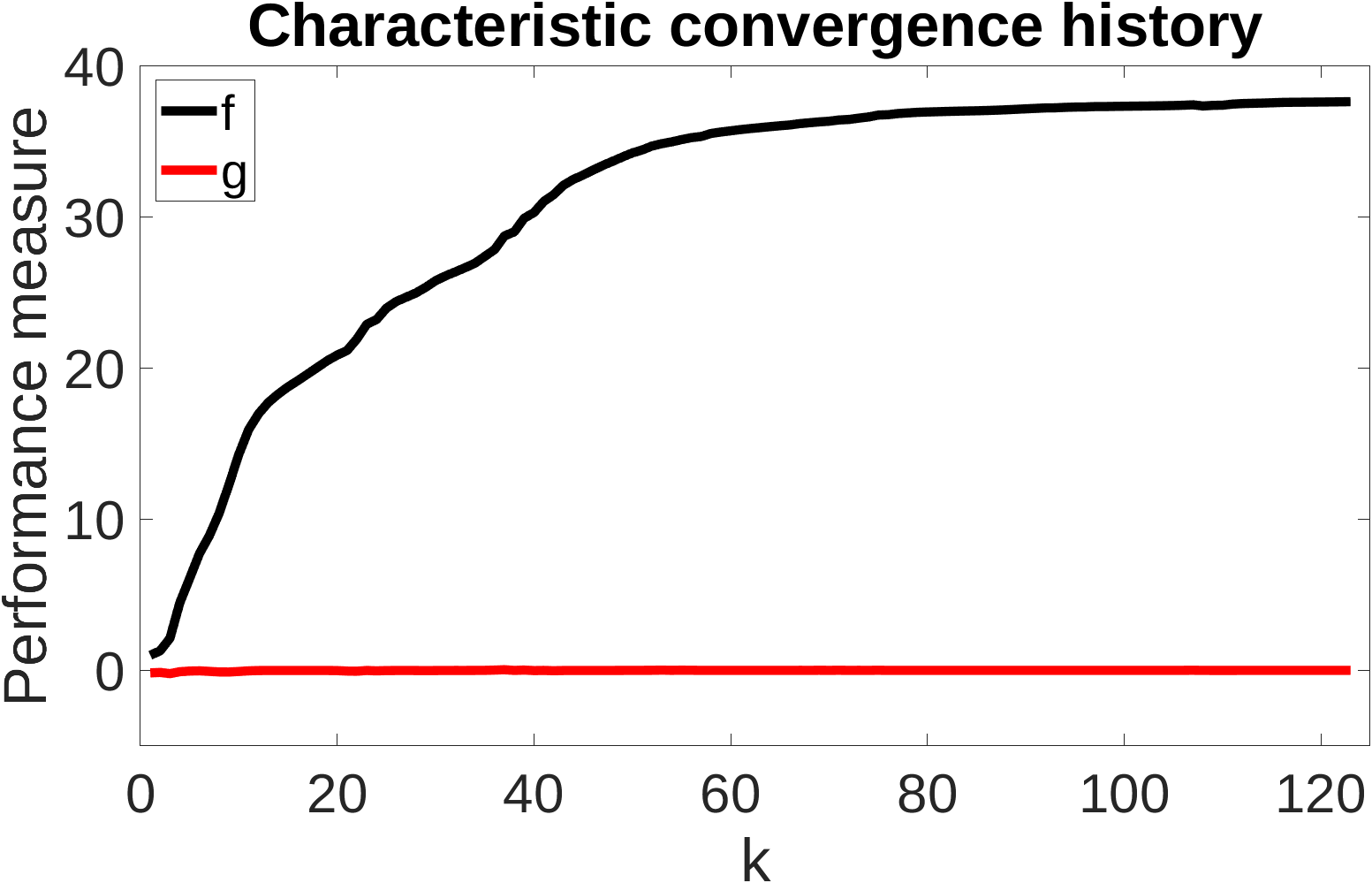}
	\caption{Characteristic convergence history of problem formulation $\mathcal{P}$: objective $f$ and constraint $g$ as a function of design iteration $k$. Note the objective is relative with respect to the first iteration, \textit{i.e.} $f^{\{0\}} = 1$. This specific convergence plot is generated by \texttt{flexure(100,100,tx,ty,1.2)}.}
	\label{fig:char_conv}
\end{figure}

Complex topologies appear for different less common design cases, such as those shown in \cref{fig:2dmaxtxminty,fig:2dmaxtxyminrz,fig:2dmaxtyrzmintx}. These results, to the best knowledge of the authors, have not been reported in literature.
Increasing the number of DOCs and/or DOFs generally results in more complex (number of bodies and rotation points) and innovative topologies, see \cref{fig:2dmaxtxyminrz,fig:2dmaxtyrzmintx}.
The convergence history of the responses for a planar optimization problem with representative set of input parameters is shown in \cref{fig:char_conv}.
For feasible input parameters, the convergence history of problem $\mathcal{P}$ is characterized by a quickly active and satisfied constraints and a steadily and smoothly increasing performance, converging within a limited number of iterations.

\begin{figure*}
	\centering
	\hfill
	\begin{subfigure}[t]{0.3\textwidth}
		\centering
		\includegraphics[width=\textwidth]{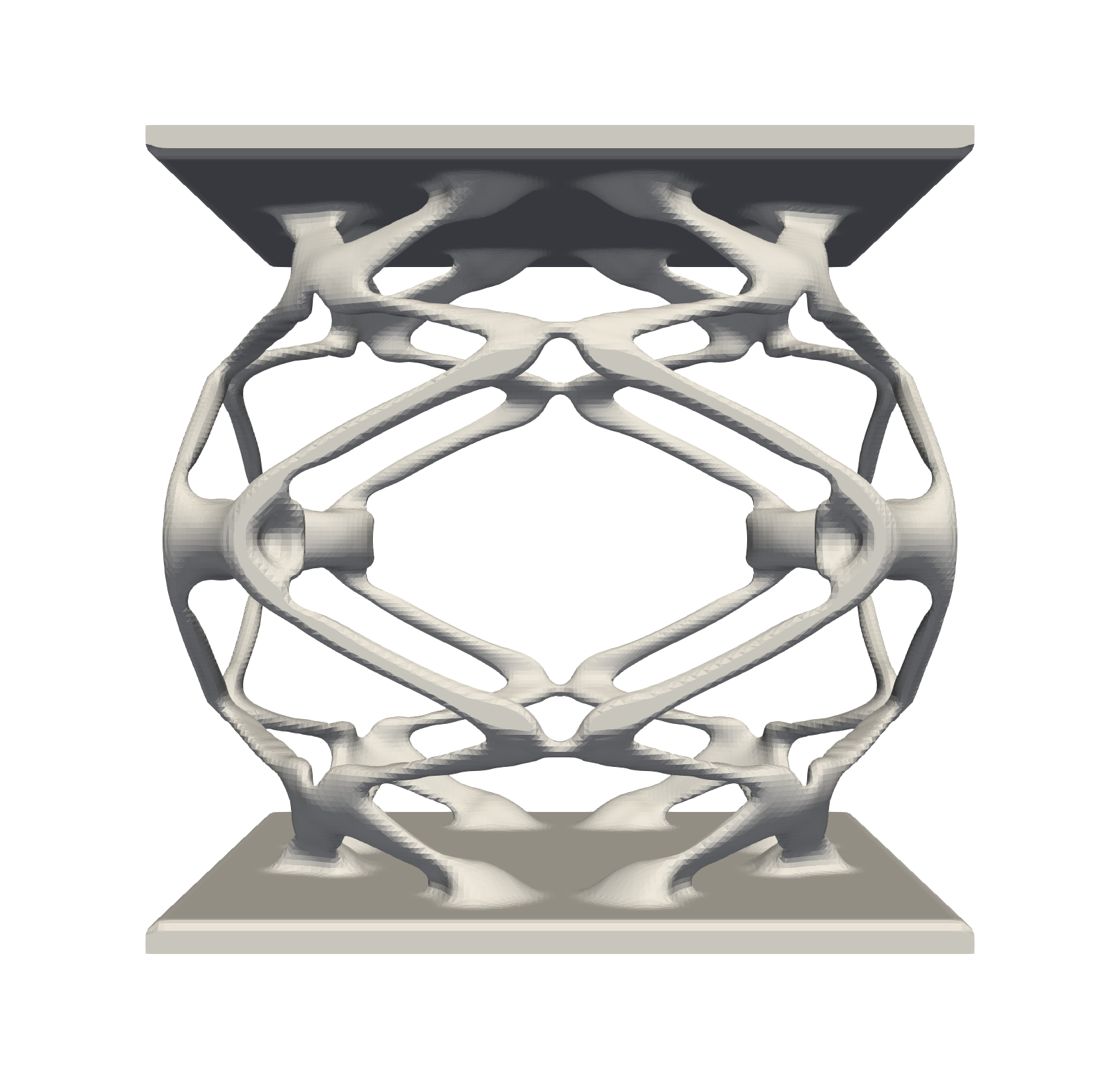}
		\caption{\texttt{doc=[rz], dof=[tx,ty,tz,rx,ry]}}
	\end{subfigure}
	\hfill
	\begin{subfigure}[t]{0.3\textwidth}
		\centering
		\includegraphics[width=\textwidth]{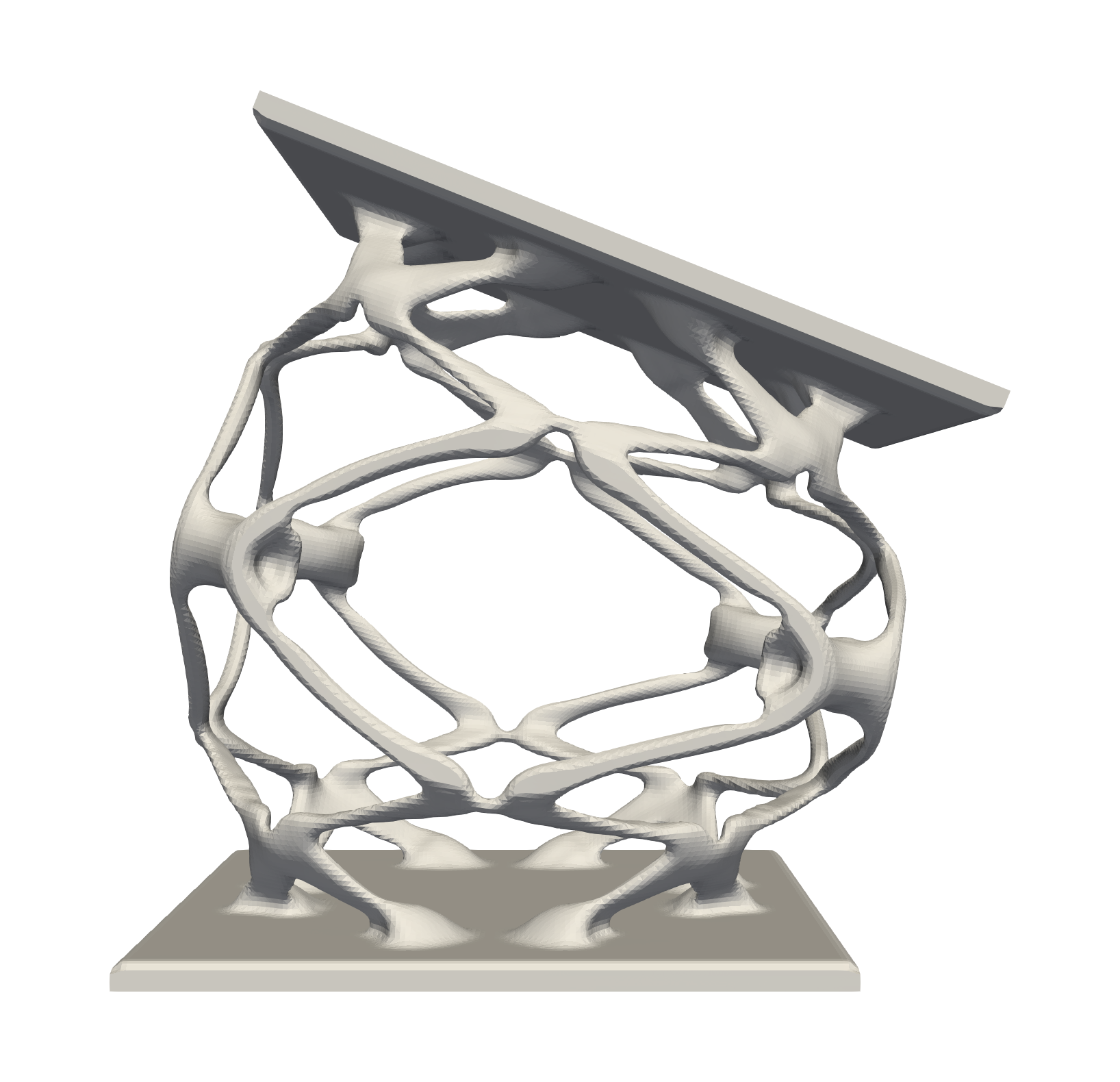}
		\caption{\texttt{ry}}
	\end{subfigure}
	\hfill
	\begin{subfigure}[t]{0.3\textwidth}
		\centering
		\includegraphics[width=\textwidth]{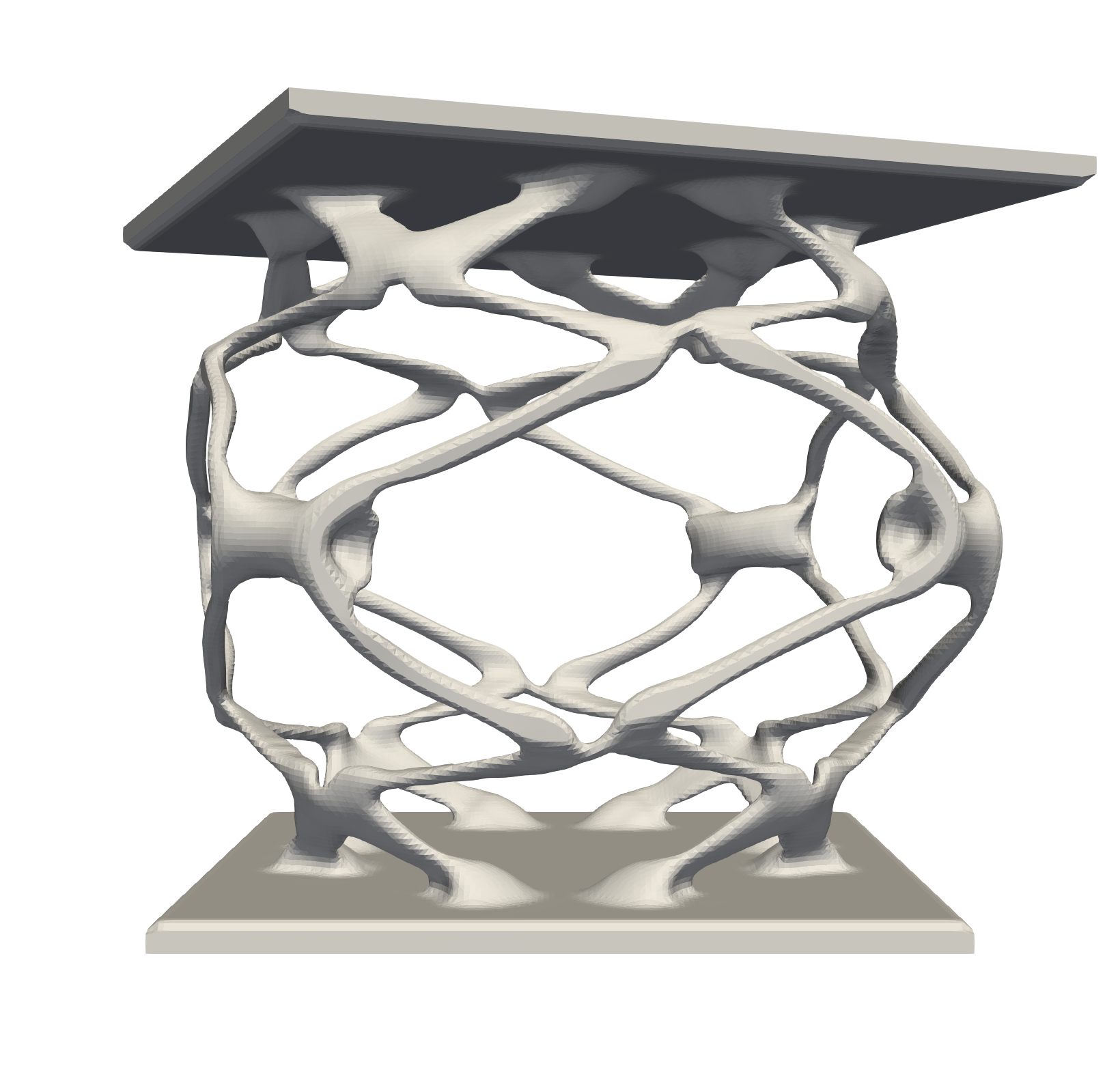}
		\caption{\texttt{rz}}
	\end{subfigure}
	\hfill
	\vskip\baselineskip
	\hfill
	\begin{subfigure}[t]{0.3\textwidth}
		\centering
		\includegraphics[width=\textwidth]{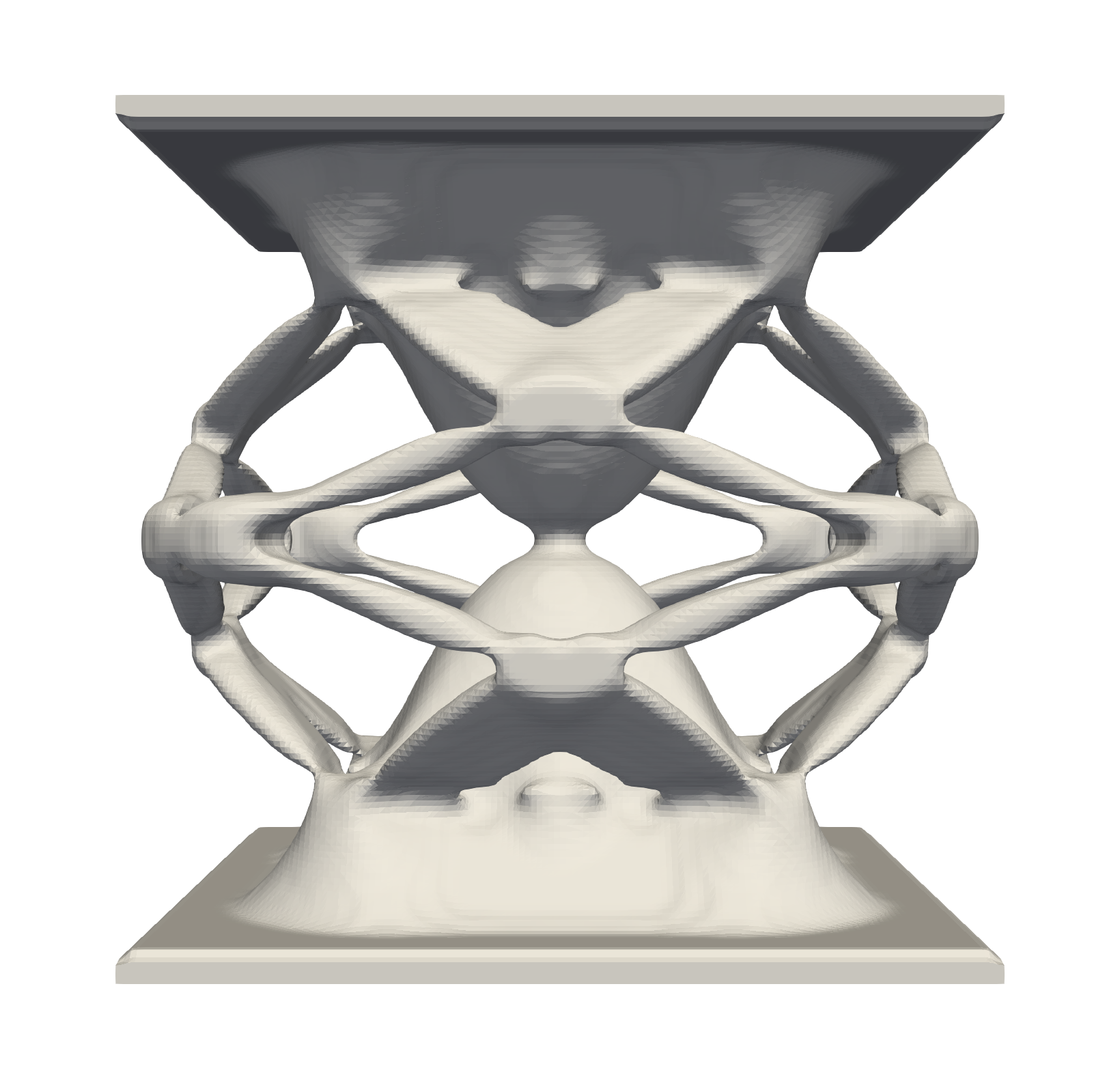}
		\caption{\texttt{doc=[tx,ty,tz,rz], dof=[rx,ry]}}
	\end{subfigure}
	\hfill
	\begin{subfigure}[t]{0.3\textwidth}
		\centering
		\includegraphics[width=\textwidth]{figs/rotflex}
		\caption{\texttt{ry}}
	\end{subfigure}
	\hfill
	\begin{subfigure}[t]{0.3\textwidth}
		\centering
		\includegraphics[width=\textwidth]{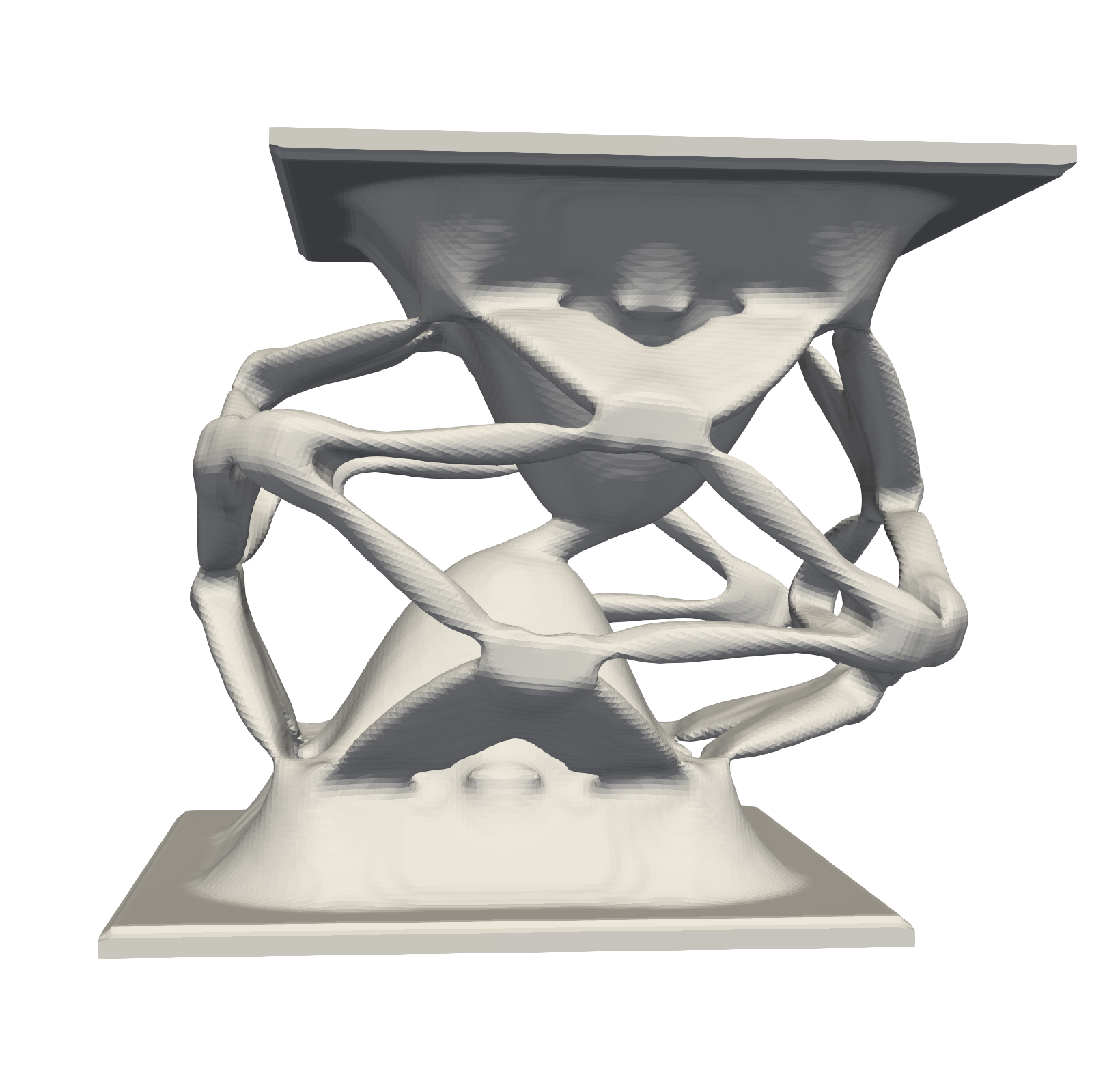}
		\caption{\texttt{tx}}
	\end{subfigure}
	\hfill
	\vskip\baselineskip
	\hfill
	\begin{subfigure}[b]{0.3\textwidth}
		\centering
		\includegraphics[width=\textwidth]{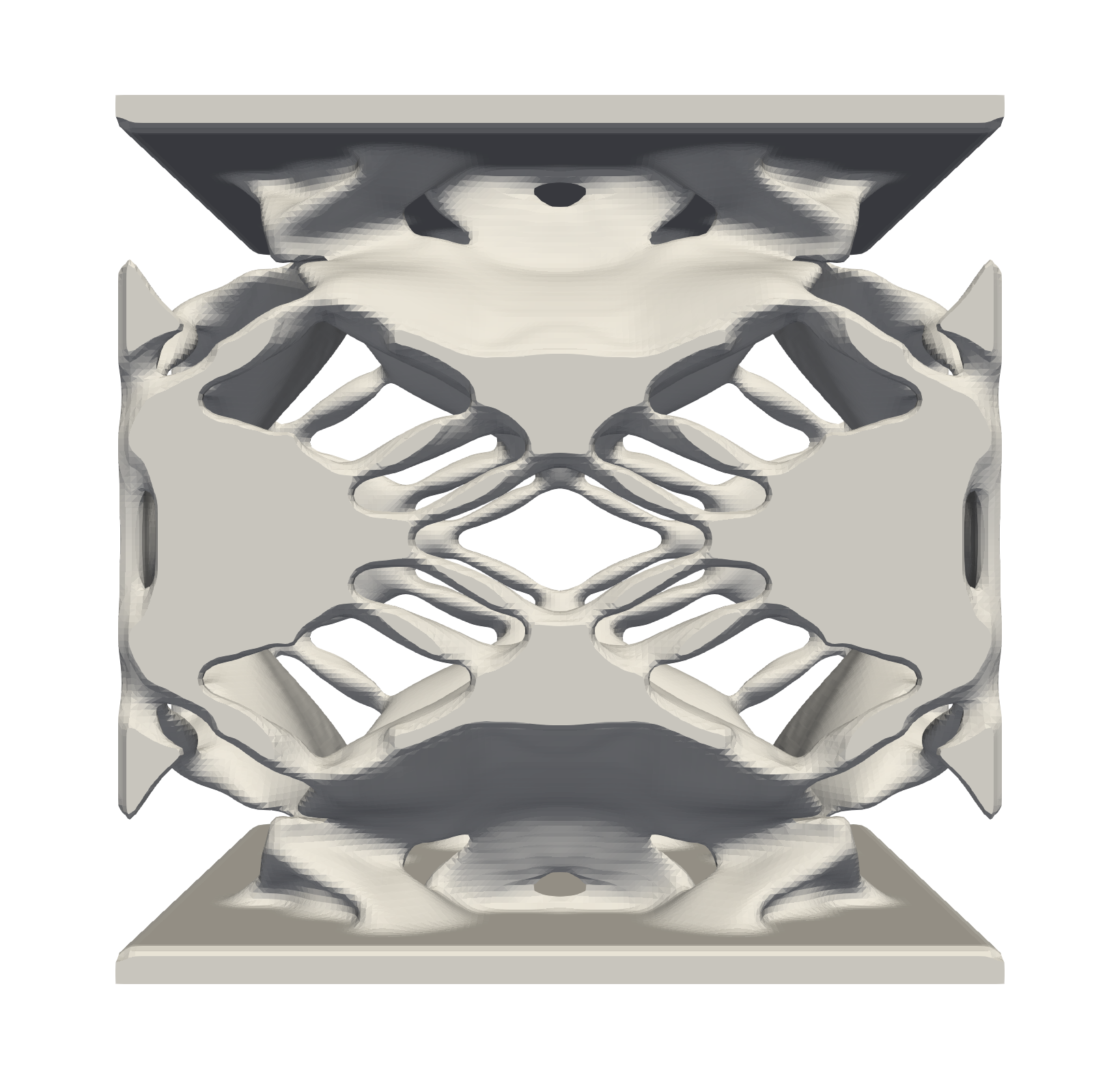}
		\caption{\texttt{ doc=[tx,ty,rx,ry,rz], dof=[tz]}}
	\end{subfigure}
	\hfill
	\begin{subfigure}[b]{0.3\textwidth}  
		\centering 
		\includegraphics[width=\textwidth]{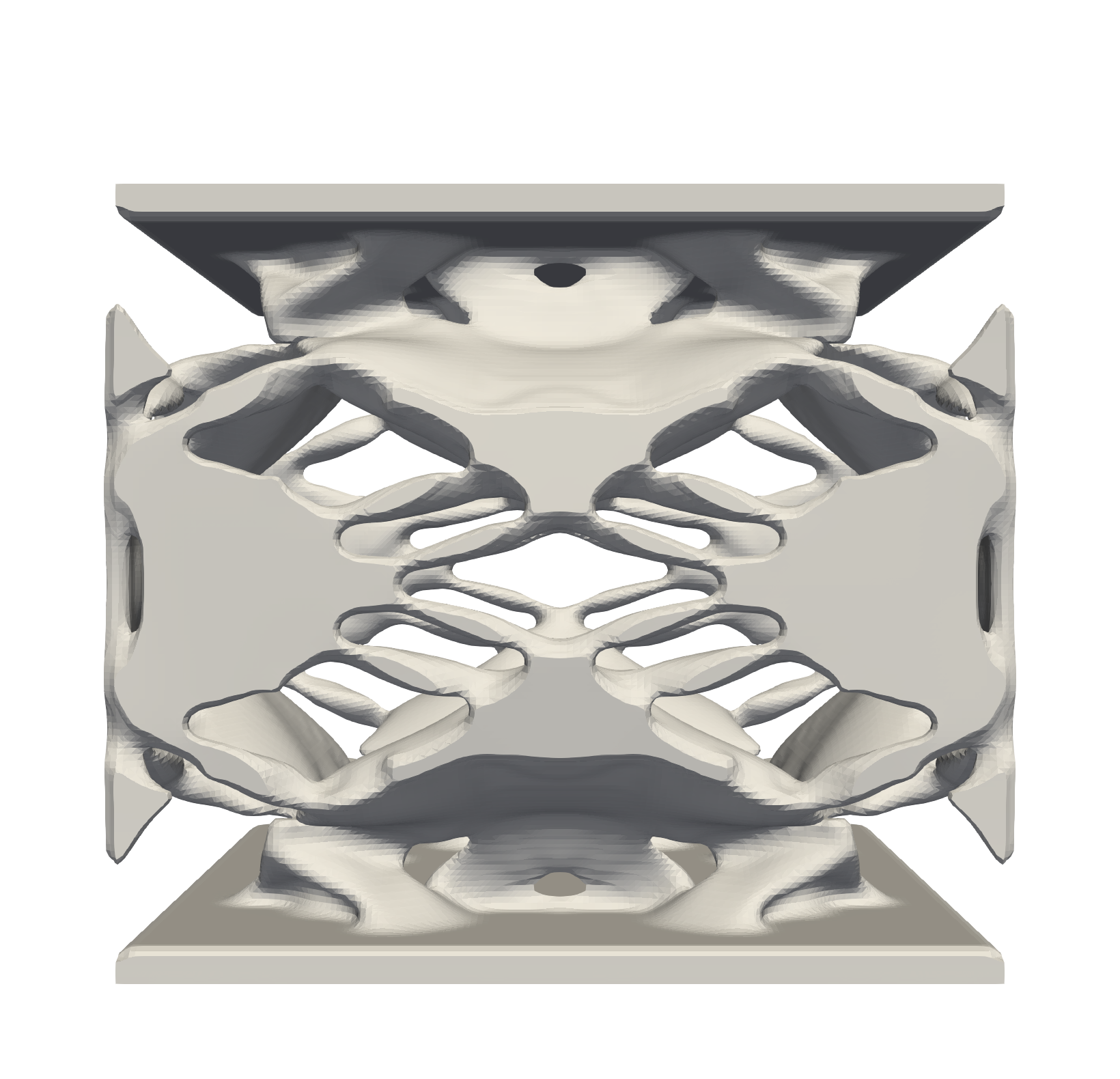}
		\caption{\texttt{tz}}
	\end{subfigure}
	\hfill
	\begin{subfigure}[b]{0.3\textwidth}   
		\centering 
		\includegraphics[width=\textwidth]{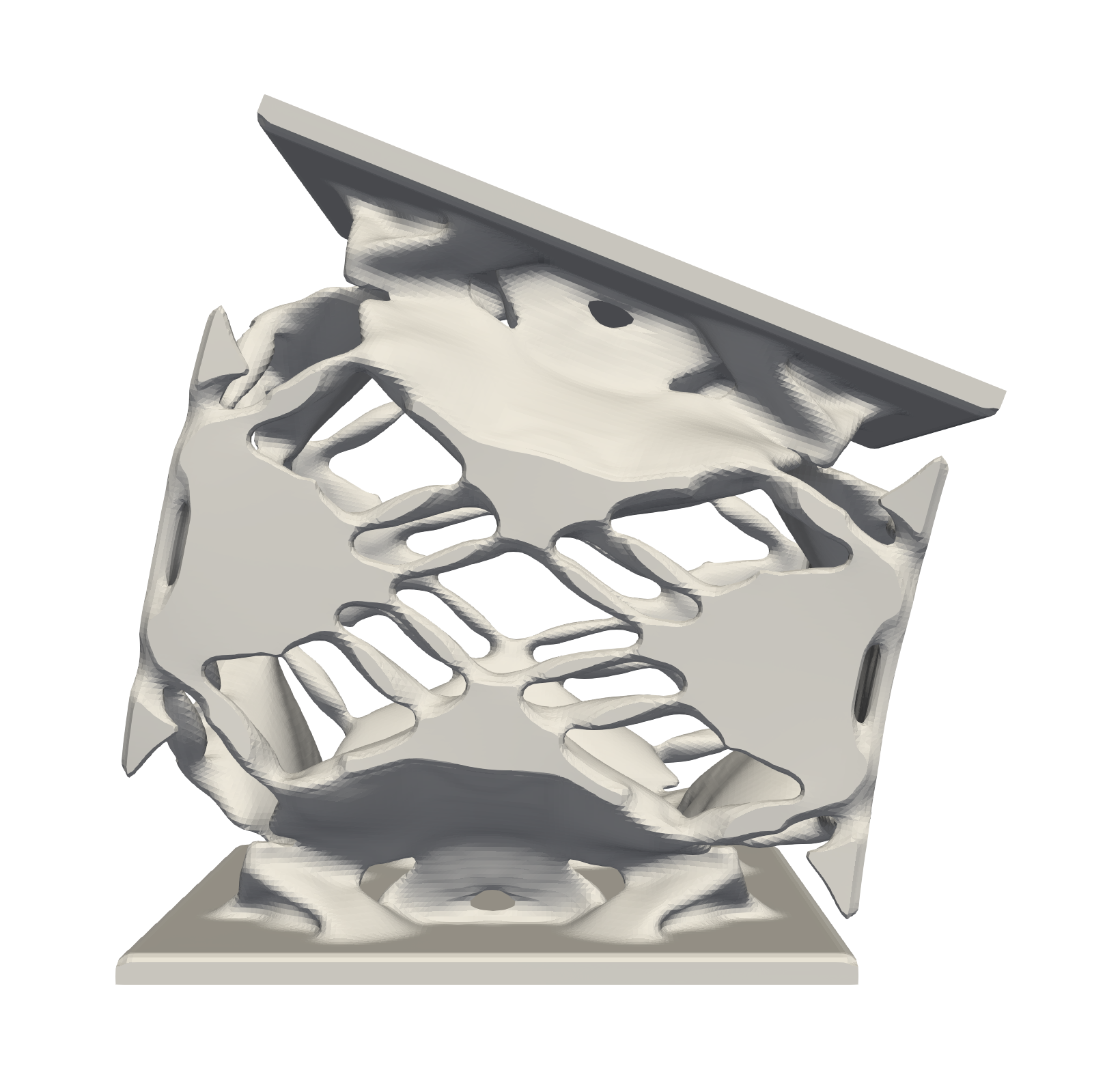}
		\caption{\texttt{ry}}
	\end{subfigure}
	\hfill
	\caption{High-resolution 3D flexure designs generated using a \texttt{C++} implementation. The number of degrees allows for a high number of variations. From left to right: the topology as a result of solving $\mathcal{P}$ for the given set of DOC(s) and DOF(s), and corresponding deformed topology for a DOF and DOC. Note that the deformations are highly scaled for visualization purposes.}
	\label{fig:flexures3d}
\end{figure*}

High-resolution 3D topologies are presented in \cref{fig:flexures3d}. Those topologies are examples of the high variety of designs that can be obtained based on the set of rigid body degrees in a 3D space.




The resulting topologies validate the correct working principle of the proposed formulation. 
In addition it shows optimized flexures have relatively small features with highly lumped strain energy, see \textit{e.g.} \cref{fig:2dmaxtyminrz}. 
As a result, those flexures have a small range of motion limited by the critical stress and their performance is sensitive to manufacturing errors.

In order to practically use the resulting designs, the maximum allowable stress as well as manufacturing uncertainties should be taken into consideration.
In the following we will show the possibilities of limiting stress levels and/or introducing manufacturing robustness in the formulation, without aiming to provide a thorough investigation of design parameters.
To this end, we use the resulting design from \cref{fig:2dmaxtxminty} as a reference.
We denote its objective by $f^0$ and corresponding maximum stress by $\overline{\sigma}^0$.

\subsection*{Fault-tolerant design}
The desired kinematics of a flexure are sensitive to both uniform and spatially varying geometric deviations. 
However, in classical deterministic topology optimization, the effect of such uncertain parameters on the performance of the structure is not taken into account. 
This may lead to a design that is very sensitive to manufacturing errors. 
As a consequence, the performance of the actual structure may be far from optimal. 
\textcite{Sigmund2009,Wang2010} propose a robust approach to topology optimization where the effect of uniform manufacturing errors is taken into account. 
Uniform erosion and dilation effects, from here on denoted by superscripts (e) and (d), are simulated by means of a projection method: the filtering of the design variable field is followed by a differentiable Heaviside projection using a high projection threshold $\eta^\text{e} = \eta + \Delta \eta$ to simulate an erosion and a low projection threshold $\eta^\text{d} = \eta - \Delta \eta$ to simulate a dilation.
An additional advantage of the robust formulation is the direct control of the minimum feature size of both solid and void.

For the robust design of flexures, only a slight difference of \cref{eq:Pbasic} is required, that is\footnote{We omit further explanation of this formulation, as arguments and implications are discussed extensively in \textcite{Wang2010}.}
\label{sec:robust}
\begin{equation}
	\label{eq:Probust}
	\mathcal{P}_{\eta} =\left\{
	\begin{aligned}
		\underset{\mathbf{x}}{\text{minimize}} && \quad & -f\left[\mathcal{E}_i\left[\mathbf{x}^\text{e}\right]\right], & i \in \mathbb{C}\\
		\text{subject to}&&&  \mathcal{E}_i\left[\mathbf{x}^\text{d}\right] \leq \overline{\mathcal{E}}_i, &i \in \mathbb{F}\\
		&&& \mathbf{x} \in \mathcal{X}
	\end{aligned}
	\right. ,
\end{equation}
with $\mathcal{E}\left[\mathbf{x}^\text{e}\right]$ and $\mathcal{E}\left[\mathbf{x}^\text{d}\right]$ strain energies based on the eroded and dilated fields, respectively. 
Since the eroded and dilated designs will always hold the maximum and minimum strain energies, respectively, the intermediate design can be excluded from the optimization formulation without compromising robustness in terms of length scale control \autocite{Lazarov2016}. 
This allows to, partially, reduce the added cost of the robust formulation.
Note that all responses still only involve self-adjoint strain energy terms.

\cref{fig:robust} shows the resulting designs of an optimization problem with filter radius $r=4$ finite elements, $\eta = 0.5$ and $\Delta\eta = 0.2$.
The robustness poses a heavy restriction on the achievable performance, as can be observed by the decrease in performance, \textit{i.e.} $f = 0.32 \times f^0$.
The hinges are clearly lengthened, thus distributing the strain energy over larger areas of the topology. 
In line with this observation, \textcite{Lazarov2018} shows it is possible to indirectly achieve stress-constrained topological design via length scale control.
Note the non-intuitive presence of protrusions along the center horizontal axis. Upon further investigation, it is observed that those do not add stiffness to the DOF, whilst contributing some (although little) stiffness to the DOC. Considering this lack of sensitivity, those are expected to be removed first upon, for example, introduction of a volume constraint. 

\subsection*{Stress-based design}
\label{sec:stress}
In order to limit the maximum stress for a given range of motion, or similarly extend the range of motion for a given maximum stress, one can simply extend the problem formulation $\mathcal{P}$ with stress constraints on the DOFs, which yields
\begin{equation}
\label{eq:Pstress}
\mathcal{P}_{\sigma} =\left\{
\begin{aligned}
\underset{\mathbf{x}}{\text{min}} && \quad & -f\left[\mathcal{E}_i\right], & i \in \mathbb{C}\\
\text{s.t.}&&&  \mathcal{E}_i\left[\mathbf{x}\right] \leq \overline{\mathcal{E}}_i, &i \in \mathbb{F}\\
&&&g_{\sigma_i}\left[\bm{\sigma}_i\right] \leq  \overline{\sigma}, & i \in \mathbb{F}\\
&&& \mathbf{x} \in \mathcal{X}
\end{aligned}
\right. ,
\end{equation}
where $\bm{\sigma}_i$ are the elemental stresses obtained by prescribing \dof $i$, and $\overline{\sigma}$ the maximum allowable stress, based on some theory of failure. To evaluate stress constraints, elemental strain energies are no longer usable.
Many different formulations of $g_\sigma$ are available \cite{Silva2019}.
Without loss of generality, we use the unified aggregation and relaxation approach as proposed by \textcite{Verbart2017}.

\begin{figure*}
	\centering
	\begin{subfigure}[t]{0.475\textwidth}
		\centering
		\includegraphics[trim={2cm, 2cm, 2cm, 2cm},clip, width=0.8\textwidth]{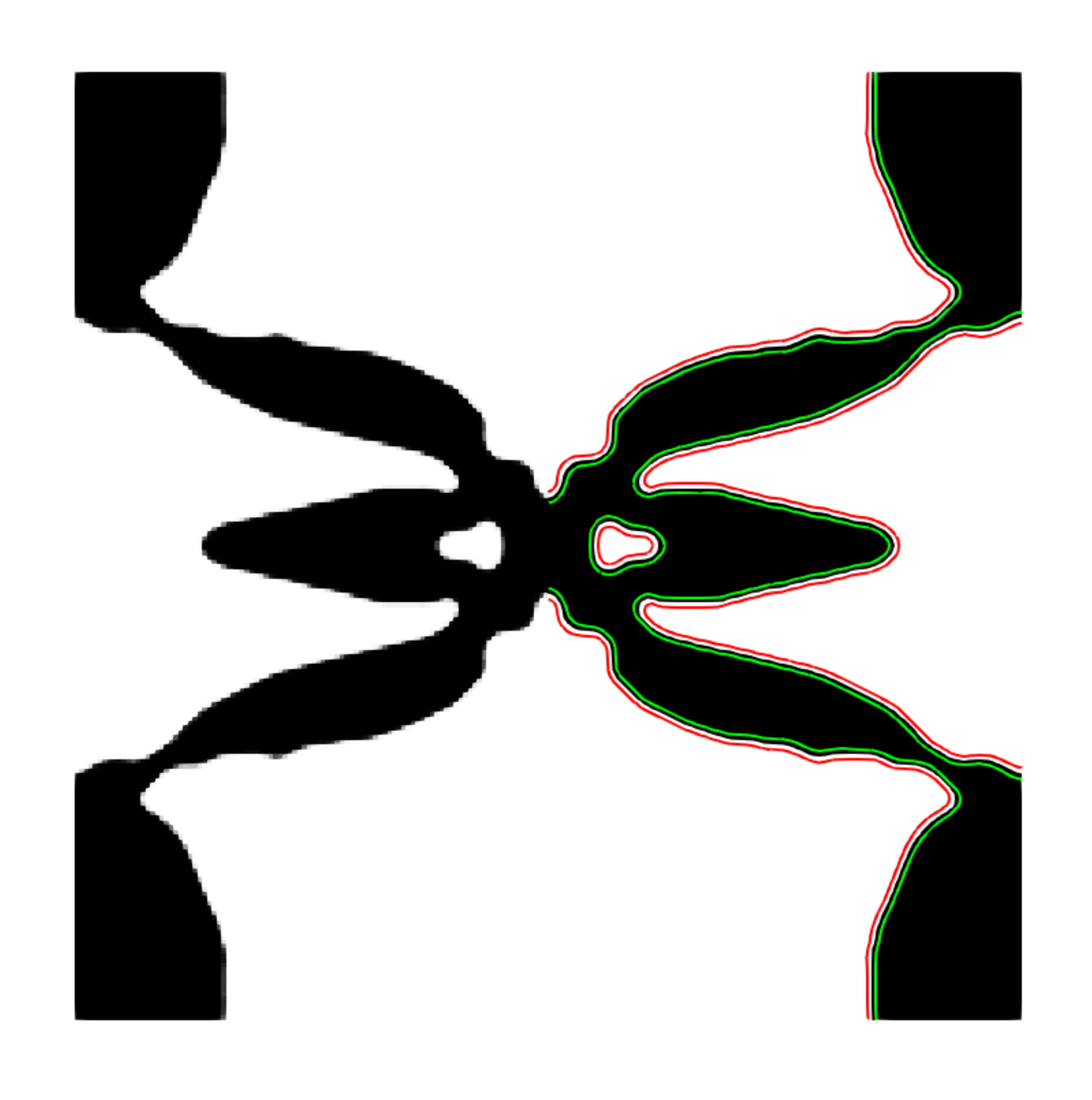}
		\caption{Design obtained by solving $\mathcal{P}_{\eta}$, see \cref{eq:Probust}. The eroded and dilated designs are indicated by, respectively, green and red contour lines. The design is robust with respect to uniform manufacturing errors \emph{and} satisfies a minimum feature size (both solid and void). However, the DOC stiffness is decreased with 68\% as compared to the design in \cref{fig:2dmaxtymintx}.}
		\label{fig:robust}
	\end{subfigure}
~~~
	\begin{subfigure}[t]{0.475\textwidth}
		\centering
	\includegraphics[trim={2cm, 2cm, 2cm, 2cm},clip, width=0.8\textwidth]{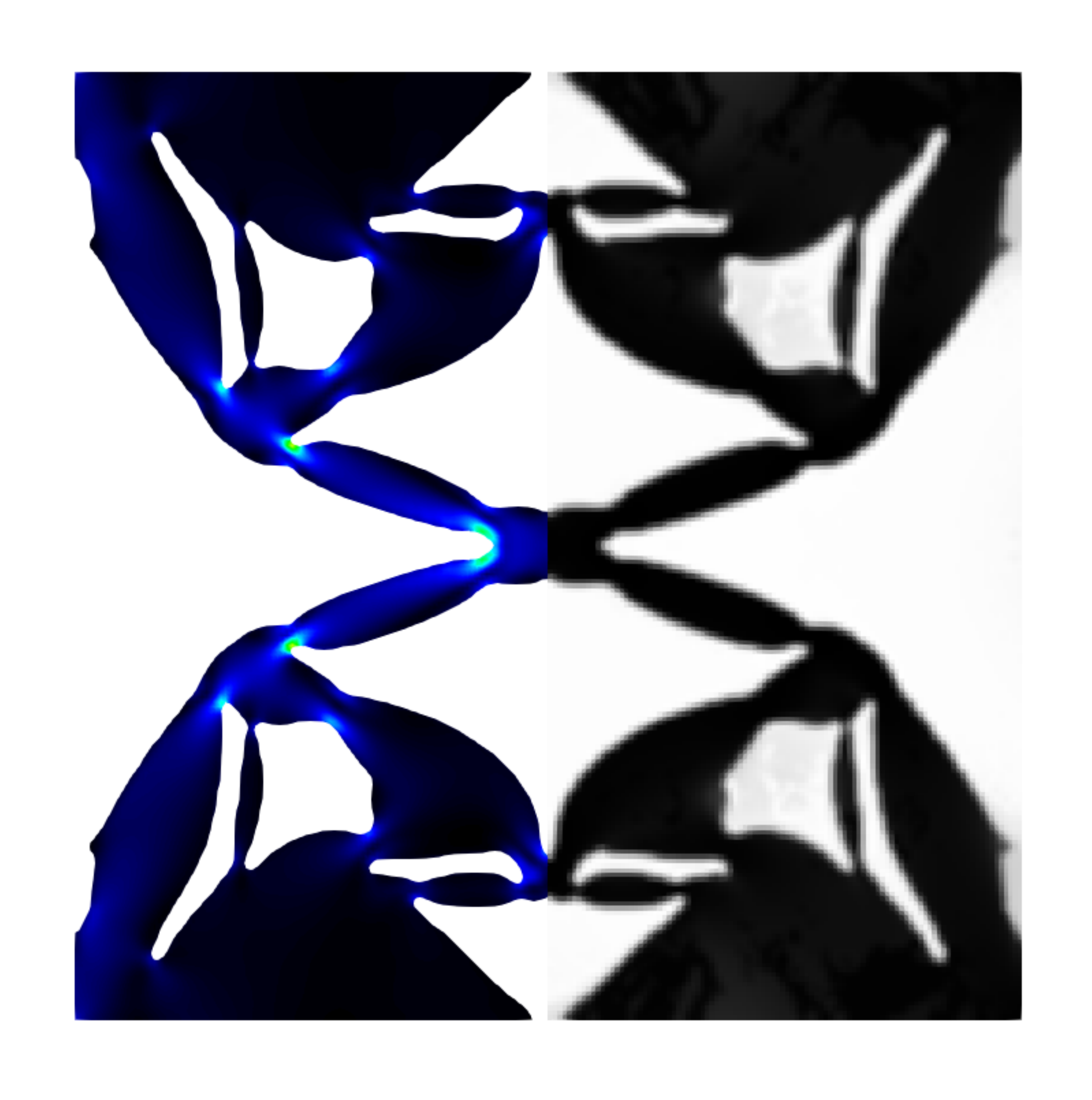}
	\caption{Design obtained by solving $\mathcal{P}_{\sigma}$, see \cref{eq:Pstress}. Major change in topology can be observed as compared to \cref{fig:2dmaxtymintx}. As a result of the applied stress constraints the range of motion is extended, with just 8\% decrease of DOC stiffness as compared to the design in \cref{fig:2dmaxtymintx}.}
	\label{fig:stress}
	\end{subfigure}

	\caption{Resulting design of the topology optimization problem from \cref{fig:2dmaxtxminty} extended with (a) the robust formulation, and (b) stress constraints.}
\end{figure*}

\cref{fig:stress} shows the resulting design of an optimization problems with $\overline{\sigma} = 0.4 \times \overline{\sigma}^0$. 
The stress constraints are satisfied by introduction of (more) hinges with a more distributed deformation energy.
Although the maximum stress is drastically reduced, the introduction of stress constraints have a relative limited impact on the performance decreases, namely $f = 0.92 \times f^0$.
This demonstrates that the proposed formulation can effectively be extended with stress constraints, yet a thorough investigation thereof is considered out of the scope of this work.

\section{Replication of results}
\label{sec:ror}
The supplementary material includes a MATLAB function (\texttt{.m}-file) that is provided to replicate the 2D single and multi-axis flexure designs from \cref{fig:flexures2d} and to use as a basis for further research.\footnote{Relevant changes or additions to the code are welcome. Contact the corresponding author for more details.}
The function file \texttt{flexure.m} is based on the \texttt{top71.m} by \textcite{Andreassen2011} and can be called using
\begin{equation*}
	\texttt{flexure(nelx,nely,doc,dof,emax)}.
\end{equation*}
Here \texttt{nelx} and \texttt{nely} define the dimensions of the rectangular design domain in terms of number of elements in \textit{x} and \textit{y}-direction. Both \texttt{doc} and \texttt{dof} are arrays containing the strings of names of \docs and \dofs, respectively. 
Parameter \texttt{emax} is an array of maximum allowable strain energies corresponding to the \dofs in \texttt{dof}. 
Further explanation on the code can be found in \cref{sec:code}.

To further encourage use of the proposed method by academics, engineers and designers and to show its versatility and ease of implementation, the material in addition contains both 2D \emph{and} 3D implementations of some more unusual geometries in commercial FEA package COMSOL Multiphysics (\texttt{.mph}-file).

Computer Aided Design (CAD) models of the 3D designs presented in \cref{fig:flexures3d} (and some more) are available in STL format (\texttt{.stl}-file) for the purpose of additive manufacturing.

\section{Discussion}
\label{sec:discussion}
Before concluding this article, a reflection on the novelty and advantages of the proposed method is in order.
Although dissimilar in formulation and implementation, the works of both \textcite{Hasse2009,Wang2009a} share the same kinetoelastic design philosophy, resulting in designs with \emph{selective compliance}. Thereto, this work can be considered both a simplification and generalization of \textcite{Hasse2009,Wang2009a}.
Also---although the proposed method does not use the separation of scales and periodicity of numerical homogenization \autocite{Bensoussan1978,Sanchez1980}---it bears some resemblance to TO of tailored materials with prescribed elastic properties by inverse homogenization \autocite{Sigmund1994b,Sigmund1995}. 
Similarly to those approaches, in the proposed formulation independent deformation patterns are prescribed (herein denoted as degrees) to optimize the structure's intrinsic properties.
The introduction of these degrees allows to easily perform design variations such as multi-axis flexures (\textit{e.g.} as demonstrated in \cref{fig:flexures2d} and \cref{fig:flexures3d}) \emph{and} straightforward adaptation to specific applications\footnote{See, \textit{e.g.}, the COMSOL design case studies in the supplementary material.}.

The simplicity and effectiveness of the formulation is directly related to the similarity with respect to the primary and most commonly used `compliance minimization' TO problem formulation by \textcite{Bendsoe1989}.
Both objective and constraints are monotonic functions of elemental strain energy, always with opposite sign of design sensitivities, which proves a well-defined optimization problem.
This results in a, relative to the state-of-the-art, easy to solve optimization problem expressed by good convergence properties.
The constraint(s) take over the `role' of the volume constraint in \textcite{Bendsoe1989} to provide auto-penalization of design variables with intermediate values, which is evidenced in binary and thereto manufacturable topologies.
The formulation requires a minimal number of independent parameters to define the optimization problem (only maximum strain energies of DOFs), simplifying its use and circumventing the common `trail-and-error' approach towards parameter value selection.

The formulation is uniquely based on strain energy measures. In contrast to the state-of-the-art, this highly simplifies implementation in/in combination with commercial FEA software packages, that generally make this data accessible for the user.
In addition, the responses are self-adjoint, consequently reducing the computational effort required to obtain design sensitivities to a minimum.
Altogether, the cost per design iteration is dominated by a single factorization/preconditioning step plus one back-substitution/iterative solve per degree. See also \cref{tab:perf} for a comparison to the state-of-the-art.

While clear advantages can be identified, the proposed formulation in not without limitations.
As presented here, it is intended for and limited to the design of short-stroke flexures, \textit{i.e.} satisfying linear strain-displacement \emph{and} stress-strain relationships. However, the prototyped samples indicate many of the resulting topologies can be used effectively in a finite range of motion.
\textcite{Duenser2021} analyzed the finite range of motion behavior for a subset of the prototyped flexures using a novel nonlinear eigenmode analysis technique. 
Results indeed indicate the flexures retain their predicted properties at least for a small finite range. As expected, for a large range of motion, stiffness properties deviate substantially.

Considering both the need for and interest in large-range compliant joints, \textit{e.g.} compliant implants, the authors pursue a study to extend the proposed formulation to design for long-stroke flexures.

It is beyond the scope of this study to include and discuss all possible variations building on this formulation, \textit{e.g.} change of geometry, degrees or objective function. Thereto we simply stress that, in our experience, the formulation does not pose any limitation to be used for shape-morphing or compliant mechanism design. As such, design studies and variations of the formulation to above mentioned applications are considered valuable future work, and the provided source code is intended to facilitate this.

\section{Conclusion}
Despite the extensive efforts devoted to research in the field of TO and the need for a effective tool to synthesize flexures for high-precision applications, a generally accepted TO problem formulation for short-stroke flexure synthesis has been absent so far.
Motivated by this, we propose a simple, versatile and computationally efficient topology optimization problem formulation.
Using mechanism degrees, this strain energy based formulation simplifies understanding and implementation of TO synthesis of flexures, and features low computation cost, smooth convergence to well-defined designs, and a minimum of tuning parameters. The base formulation is easily extended with additional design requirements and maintains its favorable properties.
Although designed for short-stroke applications, the resulting 3D designs prove practically useful within a small finite range of motion.
With source code provided to replicate the demonstrated results, this formulation is ready to be further explored and applied in academia and industry.

\section*{Declaration of interest}
The authors declare that there is no conflict of interest.

\section*{Acknowledgements}
This work was supported by the Dutch Research Council (NWO) Applied and Engineering Sciences (AES) project 16191 entitled Stable and Adjustable Mechanisms for Optical Instruments and Implants (SAMOII).

\printbibliography

\appendix

\section{Elaboration on code}
\label{sec:code}
This section elaborates on the function file \texttt{flexure.m}. The code is separated in multiple sections, starting with a section title (\textit{e.g.} \texttt{\%\% SENSITIVITY ANALYSIS}) followed by contiguous lines of code. As introduced in \cref{sec:implementation,sec:ror}, the function can be called via:
\begin{equation*}
	\texttt{flexure(nelx, nely, doc, dof, emax)}
\end{equation*}
Some possible variations are:
\begin{align*}
	\begin{aligned}
		&\texttt{flexure(100, 100, "tx", "ty", 1)}\\
		&\texttt{flexure(200, 100, ["tx","rz"], "ty", 1)}\\
		&\texttt{flexure(100, 120, "rz", ["tx","ty"], [1,0.1])}
	\end{aligned}
\end{align*}

Section \texttt{PREPROCESSING} collects and converts the user input. Degrees \texttt{tx}, \texttt{ty} and \texttt{rz} are related to displacement fields 1, 2 and 3, respectively. The section asserts the \texttt{doc} and \texttt{dof} are appropriate (no overlap, at least one DOC and DOF).

The generation of mesh (\texttt{PREPARE MESH}), preparation of FEA (\texttt{PREPARE FEA}) and preparation of density filter (\texttt{PREPARE FILTER}) are equivalent to the \texttt{top71.m} code.

Section \texttt{BOUNDARY CONDITILNS} apply the BC (prescribed displacements) for the three degrees. Instead of prescribing the nodal displacements of top and bottom nodes, we have opted to fix the nodal displacements of bottom nodes, while prescribing the nodal displacements of top nodes. In line with \cref{fig:loadcases} the displacements in x-direction (u) and y-direction (v) are found in \cref{tab:predis}.  Note that v of degree \texttt{rz} is a linear decaying function of the location in x-direction (zero in the middle). The nodal displacements are separated in \texttt{fixed} and \texttt{free} for the purpose of partitioning (see \cref{eq:eqofmotion}).
\begin{table}
	\centering
	\caption{Degree numbering and prescribed nodal displacements of the top interface.}
	\label{tab:predis}
	\begin{tabular}{c|ccc}
		\toprule
		Degree & No. & u & v\\
		\midrule
		\texttt{tx} & 1 & 1 & 0\\
		\texttt{ty} & 2 & 0 & 1\\
		\texttt{rz} & 3 & 1 & 1 - 2(x/\texttt{nelx})\\
		\bottomrule
	\end{tabular}
\end{table}

The design variable field is symmetrized in Section \texttt{FORCE SYMMETRY} to eliminate any round off errors that might, ultimately, lead to asymmetric designs. Upon commenting of the aforementioned section and using a random initial design one can explore synthesis of asymmetric designs.

The filtering of design variables (\texttt{DENSITY FILTER}), interpolation of material Young's Modulus (\texttt{MATERIAL INTERPOLATION}) and stiffness matrix assembly routine (\texttt{STIFFNESS MATRIX ASSEMBLY}) are equivalent to the \texttt{top71.m} code. Note the stiffness matrix is symmetrized (eliminate any round-off errors) to ensure the preferred solver (Cholesky factorization) is used.

Section \texttt{SOLVE SYSTEM OF EQUATIONS} contains solving the system of equations and calculation of degree strains in line with \cref{eq:y2,eq:sed}. 
To limit computational cost, the calculations are performed only on active degrees, \textit{i.e.} if a degree is not in \texttt{doc} or \texttt{dof} the corresponding nodal displacement field is not solved for.

The objective and constraint(s) are calculated in Section \texttt{RESPONSES}. Note the objective is normalized with respect to its value in first iteration, see \cref{eq:alpha}, and hence is dimensionless. 
Corresponding sensitivities (\texttt{SENSITIVITY ANALYSIS}) are in good accordance with the code in \texttt{top71.m}. If required/preferred a volume constraint can be added to the optimization problem formulation by uncommenting Section \texttt{VOLUME CONSTRAINT}.

The optimization routine provided in the code (\texttt{DESIGN OPTIMIZATION STEP}) consists of three subsequent steps, that is
\begin{itemize}
	\item determination of the variable bounds via \texttt{movelimit},
	\item generation of an approximated subproblem using the supplied \texttt{approx.m} file, and
	\item solving the strictly convex approximated subproblem using MATLAB's \href{https://nl.mathworks.com/help/optim/ug/fmincon.html}{\texttt{fmincon}}\footnote{The use of this functionality requires installation of the Optimization Toolbox.} interior-point algorithm.
\end{itemize}
The movelimit strategy detects oscillations of design variables and increases/decreases the \texttt{movelimit} accordingly. The behavior of the movelimit strategy can be adapted by changing \texttt{mlinit}, \texttt{mlincr} and \texttt{mldecr}.

For computational efficiency and improved convergence behaviour it is recommended to substitute the provided optimization routine with the original MMA by \textcite{Svanberg1987}. For simplicity the function call is added as a comment on Section \texttt{ORIGINAL MMA}.
Section \texttt{TERMINATION CRITERIA} calculates the norm of the KKT conditions as well as mean variable change, followed by Section \texttt{VARIABLE UPDATE} handling history information.

Finally, in Section \texttt{PRINT RESULTS} and \texttt{PLOT DESIGN} the performance measures (objective, constriants, mean variable change, KKT norm and strain energies) are printed to command window.

Section \texttt{TERMINATION} checks constraint feasibility as well as design variable and KKT norm change with respect to user defined tolerances. If satisfied, the optimization problem is terminated.

\end{document}